\documentclass[11pt,a4paper]{article}
\pdfoutput=1
\usepackage{jheppub}
\usepackage{rotating}
\usepackage{array}
\usepackage{amsmath}
\usepackage[normalem]{ulem}
\usepackage{slashed}
\usepackage{booktabs}
\usepackage[pdftex,table]{xcolor}
\usepackage{units}

\usepackage{multirow}

\newcommand{\vev}{v_\text{EW}}

\preprint{DESY-17-052}

\title{Dark matter self-interactions from a general spin-0 mediator}

\author{Felix Kahlhoefer,}
\author{Kai Schmidt-Hoberg}
\author{and Sebastian Wild}

\affiliation{DESY, Notkestra\ss e 85, D-22607 Hamburg, Germany}

\emailAdd{felix.kahlhoefer@desy.de}
\emailAdd{kai.schmidt-hoberg@desy.de}
\emailAdd{sebastian.wild@desy.de}

\abstract{
Dark matter particles interacting via the exchange of very light spin-0 mediators
can have large self-interaction rates and obtain their relic abundance
from thermal freeze-out. At the same time, these models face strong
bounds from direct and indirect probes of dark matter as well as a
number of constraints on the properties of the mediator. We investigate
whether these constraints can be consistent with having observable
effects from dark matter self-interactions in astrophysical systems. For
the case of a mediator with purely scalar couplings we point out the highly
relevant impact of low-threshold direct detection experiments
like CRESST-II, which essentially rule out the simplest realization of
this model. These constraints can be significantly relaxed if the
mediator has CP-violating couplings, but then the model faces strong
constraints from CMB measurements, which can only be avoided in special regions of parameter space.}


\begin{document}
\maketitle

\section{Introduction}

While the existence of dark matter (DM) has only been inferred from its gravitational interactions, additional couplings are usually required to explain its observed relic abundance. Couplings to Standard Model (SM) states are strongly constrained by the combination of direct, indirect and collider searches for the DM particle. Interactions within the dark sector, in contrast, are much less constrained and DM could have significant self-interactions affecting its behaviour on astrophysical and cosmological scales. Indeed, one of the original motivations to consider self-interacting DM (SIDM) scenarios was the realisation that long-standing small-scale structure problems encountered in the collisionless cold DM paradigm may be successfully addressed~\cite{deLaix:1995vi,Spergel:1999mh}. In order to have observable effects on astrophysical scales, the DM self-scattering cross section has to be sizeable, 
of order $\sigma / m_\psi \sim 1 \: \mathrm{cm^2\,g^{-1}}$ \cite{Buckley:2009in,Feng:2009hw,Feng:2009mn,Loeb:2010gj,Zavala:2012us,Vogelsberger:2012ku}, 
where $m_\psi$ is the DM mass. 

It is furthermore desirable that DM self-interactions be enhanced for small relative DM velocities, in order to generate effects at small scales (such as dwarf galaxies) while at the same time
avoiding the strong constraints from systems with large velocities (such as galaxy clusters)~\citep{Markevitch:2003at,Randall:2007ph,Peter:2012jh,Rocha:2012jg,Kahlhoefer:2013dca,Harvey:2015hha,Kaplinghat:2015aga}. 
Such a velocity dependence is most naturally achieved in models where a light particle $\phi$ (with $m_\phi \ll m_\psi$) mediates the DM 
interactions~\cite{Ackerman:mha,Feng:2009mn,Buckley:2009in,Feng:2009hw,Loeb:2010gj,Aarssen:2012fx,Tulin:2013teo,Kaplinghat:2015aga}. In such a set-up, a new DM annihilation channel becomes available and thermal DM freeze-out can proceed via $\chi \chi \rightarrow \phi \phi$~\cite{Pospelov:2007mp}. This way the observed DM relic abundance can be obtained even for very small couplings of the mediator to the SM. Nevertheless, the mediator ought to couple to the SM at some degree in order to establish thermal contact between the dark sector and the SM and to allow for the mediator to decay after thermal DM freeze-out so that it does not come to dominate the energy density of the Universe~\cite{Kaplinghat:2013yxa,DelNobile:2015uua,Bernal:2015ova}.\footnote{If the dark sector is much colder than the visible sector, the mediator could also be stable. Another way to evade overclosure is if the mediator decays into an even lighter dark sector state. We do not consider these possibilities further in this work.}
In fact, in many models there are strong upper bounds on the lifetime of the mediator from Big Bang Nucleosynthesis (BBN), which impose a lower bound on the coupling of the mediator to SM states.
Conversely, a tight upper bound on the coupling can be obtained from DM direct detection experiments, because the DM-nucleon scattering cross section is strongly enhanced for very light mediators.

Regarding specific realisations of this general framework, the main focus has been on vector and scalar mediators, which both give rise to a Yukawa potential in the non-relativistic limit and can therefore induce large DM self-interactions~\cite{Feng:2009hw,Tulin:2013teo,Cyr-Racine:2015ihg}. It was recently shown~\cite{Bringmann:2016din} that the case of a vector mediator, for which DM annihilation into mediators is an $s$-wave process, is strongly constrained by indirect detection experiments and Cosmic Microwave Background (CMB) measurements due to a large Sommerfeld enhancement~\cite{Sommerfeld,ArkaniHamed:2008qn} and it is not possible to obtain sizeable self-interaction cross sections. These constraints do not apply to scalar mediators, for which $s$-wave annihilation is forbidden by CP conservation.

For scalar mediators, on the other hand, there is significant tension between direct detection experiments and constraints from BBN~\cite{Kaplinghat:2013yxa,Kainulainen:2015sva}. The reason is that scalar mediators are expected to have couplings to SM fermions proportional to their mass and therefore interact much more strongly with nucleons than with electrons. It is thus difficult to achieve sufficiently small DM-nucleon scattering cross sections to satisfy bounds from direct detection experiments and at the same time ensure that the mediator decays before BBN into electrons or photons.

In this work we consider CP violation as a possible way to alleviate the tension between the various constraints for spin-0 mediators. The central observation is that both DM self-scattering and DM-nucleon scattering are substantially suppressed for pseudoscalar couplings. Allowing for different CP phases in the dark sector and in the SM sector therefore makes it possible to obtain large self-interactions while at the same time evading constraints from direct detection experiments. However, if CP is no longer conserved, DM annihilation can proceed via $s$-wave processes, reintroducing the strong constraints from indirect detection and CMB observations. 

To investigate whether all constraints and requirements can be satisfied simultaneously, we perform a detailed study of the full parameter space for a fermionic DM particle interacting via a spin-0 mediator with general CP phases. We revisit the case where the mediator has purely scalar couplings both to DM and to SM particles and point out the importance of recent results from the CRESST-II experiment~\cite{Angloher:2015ewa,Angloher:2017zkf}. This experiment is sensitive to DM masses in the GeV region and rules out the most interesting remaining parameter region. We then turn to the case where the mediator has pseudoscalar couplings to SM particles but an arbitrary CP phase in the dark sector. We show that indirect detection and CMB constraints are so strong that the CP phase must be very close to either purely scalar or purely pseudoscalar couplings. In both of these cases it is possible to obtain large self-interaction cross sections consistent with all experimental and observational constraints.

The article is structured as follows. We introduce the model in section~\ref{sec:set-up} and discuss annihilation rates, Sommerfeld enhancement and the relic density calculation. Section~\ref{sec:constraints} is dedicated to the various experimental and observational constraints on the parameter space, including a detailed discussion of the DM momentum transfer cross section. Our results are presented in section~\ref{sec:results}, first for the case of purely scalar interactions and then for arbitrary CP phases in the dark sector. We summarize and conclude in section~\ref{sec:conclusions}. Additional material is provided in the appendices~\ref{app:toymodel}--\ref{app:selfi}.

\section{General set-up}

\label{sec:set-up}

We consider a scenario in which the DM particle is a Dirac fermion $\psi$ coupled to a spin-0 particle $\phi$ with mass $m_\phi < m_\psi$. The crucial difference to most other studies of this set-up is that we do not make the assumption that the interactions of $\phi$ conserve CP (see also~\cite{Dienes:2013xya,Dolan:2014ska,Kainulainen:2015sva,Ma:2016tpf,Arina:2017sng}). Its interactions with DM can therefore be written as
\begin{equation}
\mathcal{L}_\text{DM} \supset - \,  y_\psi \cos \delta_\psi \, \, \bar \psi \psi \phi \, - \, i \, y_\psi \sin \delta_\psi \, \bar \psi \gamma^5 \psi \phi \; .
\label{eq:LDM}
\end{equation}
This coupling structure is equivalent to $y_\psi e^{i \delta_\psi} \, \bar{\psi}_R \psi_L \, \phi + \, \text{h.c.}$, so $\delta_\psi$ parametrizes the CP-violating phase. Note that for $\delta_\psi = 0$ ($\delta_\psi = \pi/2$) the Lagrangian conserves CP provided $\phi$ is an even (odd) eigenstate under CP. In this case the Lagrangian reduces to the well-studied cases of DM interacting via a scalar or pseudoscalar mediator, respectively.

In addition to mediating the interactions of DM particles with each other, the $\phi$ boson can also couple to SM fermions. These interactions ensure that $\phi$ is unstable and furthermore provide a mechanism for bringing the DM particle into thermal equilibrium with the SM states. In order not to induce unacceptably large effects in precision measurements of SM flavour observables, we require that the couplings of $\phi$ to SM states are consistent with the hypothesis of minimal flavour violation~\cite{Abdallah:2015ter}, which implies that the couplings of $\phi$ to the various SM fermions $f$ should be proportional to their respective masses $m_f$. Nevertheless, we do allow for CP violation also in the visible sector, i.e.\ we consider the following interaction Lagrangian:
\begin{equation}
\mathcal{L}_\text{mixing} \, = \, - \, y_{\text{SM}} \sum_f \left[ \frac{m_f}{\vev} \cos \delta_{\text{SM}} \, \, \bar f f \phi \, + \, i \frac{m_f}{\vev} \sin \delta_{\text{SM}} \, \, \bar f \gamma^5 f \phi \right] \,.
\label{eq:Lmix}
\end{equation}
Note that this interaction is only apparently renormalisable because it can only be induced after electroweak symmetry breaking~\cite{Bell:2015sza}. The parameter $y_\text{SM}$ must therefore be proportional to the electroweak vacuum expectation value $\vev$ divided by the scale $\Lambda$ of some (unspecified) high energy mechanism. We will be interested only in the case where $y_\text{SM} \ll 1$, so that it is fully sufficient to consider the effective low-energy description given in eq.~(\ref{eq:Lmix}).

In the following, we will adopt a phenomenological point of view, meaning that we remain mostly agnostic about the origin of the CP-violating phases in eqs.~(\ref{eq:LDM}) and (\ref{eq:Lmix}) and analyse the model by treating these phases as free parameters. Nevertheless, there are strong constraints on new sources of CP violation in the SM, for example from measured upper bounds on the electric dipole moments of light leptons~\cite{Chen:2015vqy,Marciano:2016yhf} and nuclei~\cite{Mantry:2014zsa}. These constraints are evaded as long as $y_\text{SM}$ is sufficiently small and $\delta_\text{SM}$ is close to either $0$ or $\pi/2$. To demonstrate that such a configuration can occur quite naturally, we discuss a toy model of spontaneous breaking of CP symmetry in appendix~\ref{app:toymodel}. In this model, one finds that $\delta_\psi$ can take any value between $0$ and $\pi/2$, while $\delta_\text{SM}$ is always very close to $\pi/2$  as a natural consequence of $y_\text{SM} \ll y_\psi$.

\subsection{DM annihilation processes}

In contrast to $y_\text{SM}$, the coupling $y_\psi$ between the DM particle and $\phi$ can be large. Since we are interested in the case where $\phi$ is lighter than $\psi$, the cross section for a pair of DM particles to annihilate into $\phi$ bosons can therefore be sizeable. In the limit $m_\phi \ll m_\psi$ we obtain:\footnote{We note that in the limit of a scalar mediator ($\delta_\psi = 0$) we obtain an expression that is a factor 2 smaller than the corresponding one in~\cite{Tulin:2013teo}. Our expression does however agree with~\cite{Baldes:2017gzw} in this limit.}   
\begin{align}
(\sigma v)_{\psi \bar \psi \rightarrow \phi \phi} \, \simeq \, \frac{y_\psi^4 \sin^2(2 \delta_\psi)}{32 \pi m_\psi^2} + \frac{y_\psi^4 \left[3 + 8 \cos(2 \delta_\psi) + 7 \cos(4 \delta_\psi)\right]}{768 \pi m_\psi^2} \cdot v^2 \; ,
\label{eq:2to2}
\end{align}
where we have only kept the first two terms in an expansion in the relative velocity of the two DM particles $v$. We note that in the CP-conserving case, i.e. for $\delta_\psi = 0$ or $\delta_\psi = \pi/2$, the velocity-independent ($s$-wave) contribution vanishes. This is a consequence of the fact that in this case the annihilation process must proceed with non-zero angular momentum to ensure that initial and final state have the same CP quantum numbers. This argument does not apply for other values of $\delta_\psi$, so that the annihilation cross section for $\psi \bar \psi \rightarrow \phi \phi$ is in general non-zero in the limit $v\rightarrow 0$. 

Another process of interest is the annihilation into three (very light) $\phi$ bosons:
\begin{align}
(\sigma v)_{\psi \bar \psi \rightarrow \phi \phi \phi} \, \simeq \, \frac{(7 \pi^2-60) y_\psi^6 \sin^6 \delta_\psi}{1536 \pi^3 m_\psi^2} \,.
\label{eq:2to3}
\end{align}
Despite the additional phase space and coupling suppression, this process turns out to contribute significantly to the total annihilation cross section in certain regions of the parameter space, as it is $s$-wave even in the case of pure pseudoscalar couplings.\footnote{For values of $\delta_\psi$ different from $0$ or $\pi/2$ this process exhibits an infrared divergence. This can be attributed to the initial state radiation of a massless scalar, which can lead to an on-shell intermediate DM particle. However, as the $2 \rightarrow 3$ annihilation is expected to be only relevant for $\delta_\psi \simeq \pi/2$ (as in this region of the parameter space the lowest-order annihilation into a pair of mediators is suppressed), it is sufficient to only take into account the contribution arising from the pseudoscalar coupling $y_\psi \sin \delta_\psi$, leading to the infrared-finite expression given in eq.~(\ref{eq:2to3}).}

If the mediator $\phi$ is sufficiently light and sufficiently strongly coupled, the cross sections given above must be multiplied with a Sommerfeld enhancement factor~\cite{Sommerfeld}, which reflects the modification of the initial-state wave function due to multiple mediator exchange: $(\sigma v)_\text{enh} = S \times \sigma v$. Approximating the Yukawa potential by a Hulth\'en potential, one finds for an $s$-wave annihilation process~\cite{Cassel:2009wt,Iengo:2009ni,Slatyer:2009vg} \begin{equation}
 S_s = \frac{\pi}{a} \frac{\sinh (2 \pi \, a \, c)}{\cosh (2 \pi \, a \, c) - \cos (2 \pi \sqrt{c - a^2 c^2})} \; ,
\end{equation}
where $a = v/(2 \alpha_\psi)$ and  $c = 6 \, \alpha_\psi \, m_\psi / (\pi^2 m_\phi)$ with $\alpha_\psi = y_\psi^2 \, \cos^2 \delta_\psi / (4\pi)$.\footnote{Note that we only use the scalar part of the coupling to calculate the Sommerfeld enhancement. We will return to this issue in more detail in the context of DM self-interactions in section~\ref{sec:constraints} and in particular in appendix~\ref{app:pseudoSIDM}.} The corresponding expression for a $p$-wave process is
\begin{equation}
 S_p = \frac{(c-1)^2 + 4 \, a^2 c^2}{1 + 4 \, a^2 c^2} \times S_s \; . 
\end{equation}
Crucially, the Sommerfeld factor depends on the relative DM velocity $v$. For $v \gtrsim \alpha_\psi$ one obtains $S_{s,p} \approx 1$, whereas for smaller velocities $S$ increases proportionally to $1/v$ in the $s$-wave case and $1/v^3$ in the $p$-wave case, so that effectively the annihilation cross section in both cases increases proportionally to $1/v$ with decreasing velocity. The Sommerfeld enhancement saturates for $v \lesssim m_\phi / (2 m_\psi)$, so the ratio of the two masses determines the maximum possible enhancement. Note that, if the model parameters are close to a resonance, the enhancement can be even larger and saturate even later.

The annihilation processes discussed above lead to two important effects. First of all, assuming that the dark sector is initially in thermal equilibrium with the SM sector, DM particles can obtain their relic abundance from thermal freeze-out in such a way that the DM relic abundance is determined by its annihilation cross section into light $\phi$ bosons. Second, since $\phi$ can decay into SM final states, DM annihilation processes can potentially be observed indirectly. For example, DM particles annihilating at the time of recombination can inject electromagnetic energy into the plasma and thereby spoil the successful predictions of the CMB radiation~\cite{Adams:1998nr,Chen:2003gz,Slatyer:2009yq,Galli:2009zc,Cline:2013fm,Liu:2016cnk}. We will first discuss the calculation of the DM relic abundance and then return to constraints from indirect measurements in the following section.

\subsection{Thermal freeze-out}
\label{sec:relic}

The DM relic abundance depends on the thermally-averaged annihilation cross section $\langle \sigma v \rangle$ in such a way that larger cross sections correspond to smaller relic abundances. The simplest way to calculate the freeze-out prediction is to compare $\langle \sigma v \rangle$ to the so-called thermal cross section $\langle \sigma v \rangle_\text{thermal}$, which yields the observed DM abundance \mbox{$\Omega_\text{DM} h^2 = 0.12$}~\cite{Ade:2015xua}:
\begin{equation}
 \Omega_\psi h^2 = 0.12 \times \frac{\langle \sigma v \rangle_\text{thermal}}{\langle \sigma v \rangle} \; .
\label{eq:omega}
\end{equation}
The thermal cross section for a Dirac DM particle is approximately given by $\langle \sigma v \rangle_\text{thermal} \approx 4.4 \cdot 10^{-26} \,\mathrm{cm^3 \, s^{-1}}$ with a slight dependence on $m_\psi$, which we take from the detailed calculation in~\cite{Steigman:2012nb}.

A more accurate analysis of the underlying Boltzmann equation reveals that the quantity that enters in the denominator of eq.~(\ref{eq:omega}) is not exactly $\langle \sigma v \rangle$ but actually
\begin{equation}
x_\mathrm{f} \int_{x_\mathrm{f}}^{\infty} \frac{\langle \sigma v \rangle}{x^2} \, \mathrm{d}x \; ,
\label{eq:xfint}
\end{equation}
where $x = m_\chi / T$ parametrizes the temperature and its value at freeze-out lies in the range $20 < x_\mathrm{f} < 30$. For example, if the annihilation cross section can be written as $\sigma v = a + b v^2$, one finds $\langle \sigma v \rangle = a + 6b/x$ and hence the denominator in eq.~(\ref{eq:omega}) should read $a + 3 b / x_\mathrm{f}$.

In our case, however, $\sigma v$ has a more complicated dependence on the velocity due to the onset of Sommerfeld enhancement for small velocities. Assuming $\sigma v = a'/v + b'/v$, the thermal averaging yields $\langle \sigma v \rangle = (a' + b') (\pi / x)^{-1/2}$. In practise, however, Sommerfeld effects are only beginning to become important during thermal freeze-out, so it is not in general clear which of the two expressions to use. We therefore adopt an approach that interpolates between the two cases and take
\begin{equation}
 \langle \sigma v \rangle \approx (\sigma v)_\text{enh}\rvert_{v = (\pi / x_\text{f})^{1/2}} \; .
\label{eq:thermal}
\end{equation}
This yields the correct result in the case that $(\sigma v)_\text{enh} \propto 1/v$, whereas for $S = 1$ one obtains $\langle \sigma v \rangle = a + \pi b / x_\text{f}$, which differs only slightly from the correct expression $a + 3 b / x_\text{f}$ obtained analytically. We find this approximation to be sufficiently accurate given the uncertainty in $x_\text{f}$ entering eq.~(\ref{eq:omega}).

To calculate the relic abundance of $\psi$ we take the values of $x_\text{f}$ provided in~\cite{Steigman:2012nb} for the $s$-wave case. This treatment ensures that our results are exactly correct in the case that freeze-out proceeds dominantly via $s$-wave processes and that Sommerfeld enhancement is not relevant, which are both good approximations in most of the parameter space that we consider. By applying our approach to the well-studied case of a vector mediator (see e.g.~\cite{Tulin:2013teo}), we confirm that our approximation is still very good in the case of $s$-wave annihilation with sizeable Sommerfeld enhancement. Furthermore, for dominantly $p$-wave annihilation, as for example in the case that $\delta_\psi = 0$, we have checked that our treatment agrees with more accurate studies in the literature including Sommerfeld enhancement~\cite{Tulin:2013teo} once we account for the uncertainty in $x_\text{f}$ inherent in the analytical approach.

\begin{figure}[t]
\center
\hspace{-0.3cm}
\includegraphics[scale=0.92]{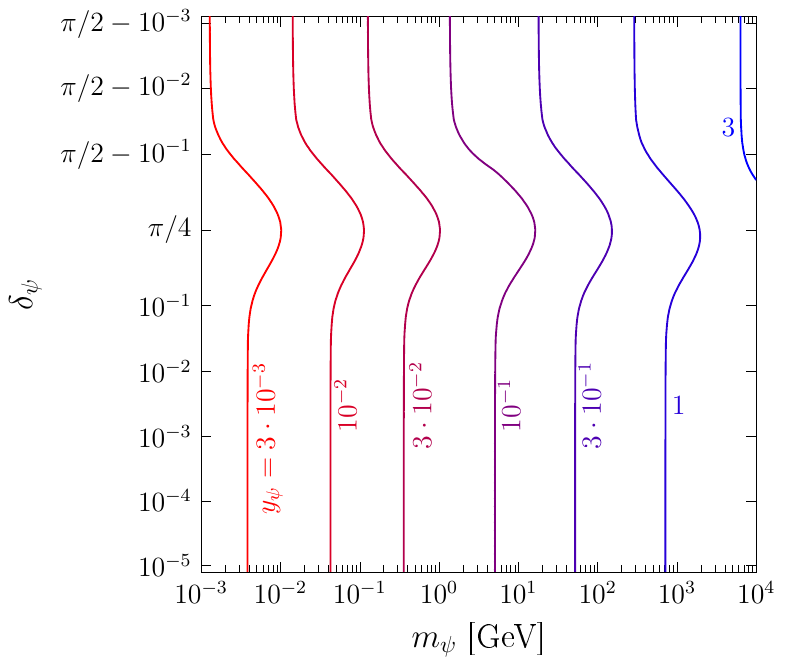}\hspace{0.2cm}
\includegraphics[scale=0.92]{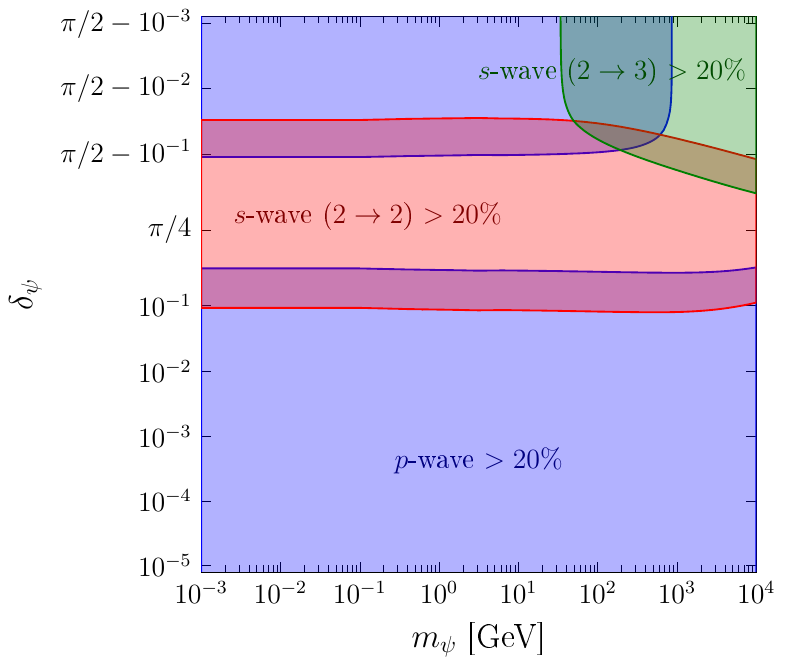}
\caption{Values of $y_\psi$ that give the correct relic abundance (left) and the relative importance of the different annihilation contributions (right).}
\label{fig:relic_density}
\end{figure}

We can therefore now proceed and calculate the thermal abundance of the DM particle $\psi$ using eq.~(\ref{eq:omega}) with $\langle \sigma v \rangle$ given in eq.~(\ref{eq:thermal}). Since the annihilation cross section depends monotonically on the coupling $y_\psi$, we can determine the value of $y_\psi$ that yields $\Omega h^2 = 0.12$ for each combination of $m_\psi$ and $m_\phi$. Note that this value is completely independent of how quickly $\phi$ decays into SM particles, and it is also independent of $m_\phi$ as long as $m_\phi \ll m_\psi$. The values obtained in this way are shown in the left panel of figure~\ref{fig:relic_density}. One observes that larger couplings are required for larger $m_\psi$, as well as for $\delta_\psi$ close to either 0 or $\pi/2$. The right panel of figure~\ref{fig:relic_density} indicates in which regions of parameter space freeze-out is dominated by $s$-wave processes and where $p$-wave processes give the largest contribution. One can furthermore see the $2 \to 3$ process becoming relevant for large $m_\psi$ and $\delta_\psi \approx \pi / 2$. From now on we will always fix $y_\psi$ to the value that reproduces the observed relic abundance and discuss the phenomenology of the remaining parameter space.

To conclude this section, let us add a number of comments on the relic density calculation. First of all, we note that for heavy DM masses and large couplings $y_\psi$ the relic abundance calculation is modified by bound-state formation, as discussed in~\cite{vonHarling:2014kha,Cirelli:2016rnw,Baldes:2017gzw} for the case of scalar mediators and in~\cite{Baldes:2017gzw,An:2016kie} for vector mediators. Since these effects are small for the parameter region that we are interested in, we do not discuss them further.

We also note that very close to a resonance there can be a second period of DM annihilation after kinetic decoupling~\cite{Dent:2009bv,Zavala:2009mi,Feng:2010zp,vandenAarssen:2012ag}, which can significantly affect the evolution of the DM abundance. As the kinetic decoupling temperature typically is in the MeV region \cite{Feng:2010zp}, reannihilations are expected to occur around keV temperatures if the annihilation cross section scales as $1/v^2$ as is the case exactly on a resonance. In this case smaller dark sector couplings are sufficient to obtain the correct relic abundance, alleviating to some degree the constraints from direct and indirect detection. We do not consider detailed properties of these very tuned regions of parameter space further in this work.

Lastly, the expansion of $(\sigma v)$ in powers of $v^2$ given in eq.~(\ref{eq:2to2}) breaks down for $m_\phi \simeq m_\psi$, requiring a more careful treatment of the dark matter freeze-out process~\cite{Gondolo:1990dk}. However, as we will see below, in this case the dark matter self-interactions are in any case negligible. Hence, in the following we restrict ourselves to $m_\phi \lesssim 0.2 m_\psi$ which covers all the parameter space relevant to our study.

\section{Experimental and observational constraints}
\label{sec:constraints}

We begin this section by reviewing the phenomenology of DM self-interactions in the context of the scenario that we consider. We will identify those regions of parameter space where self-interactions are so large that they are disfavoured by astrophysical observations as well as those regions where SIDM potentially provides a better explanation of astrophysical data than collisionless DM. We then continue by reviewing the various relevant experimental constraints on our model. These constraints can roughly be divided into constraints that are independent of the SM coupling $y_\text{SM}$, constraints that are independent of the DM coupling $y_\psi$ and constraints that depend on both couplings. The first type of constraints comprises indirect detection experiments and CMB constraints while the second one consists of astrophysical constraints and constraints from searches for rare processes. Into the third category fall direct detection experiments and the requirement of thermalisation between the two sectors.

\subsection{Dark matter self-interactions}
\label{sec:SIDM}

DM particles interacting with each other via the exchange of a very light mediator may experience large rates of self-scattering. The resulting redistribution of momentum can reduce the central densities of DM halos (thus transforming cusps into cores), reduce halo ellipticity and even lead to the evaporation of sub-halos. The impact of DM self-interactions on astrophysical objects is quantified by the momentum transfer cross section $\sigma_\mathrm{T}$, which is defined as\footnote{This expression generalizes equation (A8) from~\cite{Kahlhoefer:2013dca} to the case of particle anti-particle scattering, which was recently discussed in~\cite{Agrawal:2016quu}.}
\begin{equation}
 \sigma_\mathrm{T} = 2 \pi \int_{-1}^1 \frac{\mathrm{d}\sigma}{\mathrm{d}\Omega} (1 - |\cos \theta|) \, \mathrm{d} \cos \theta \;.
\label{eq:sigT}
\end{equation}

In a similar way as the annihilation processes discussed above, this quantity receives significant contributions from non-perturbative effects. These effects can be taken into account by solving the non-relativistic Schr\"odinger equation corresponding to the potential induced by the exchange of the mediator $\phi$, which is given by the Fourier transformation of the matrix element $\mathcal{M}(\vec{q})$ for the scattering process:
\begin{equation}
 V(r) = -\int \frac{\text{d}^3 q}{(2 \pi)^3} \, e^{i \vec{q} \cdot \vec{r}} \frac{\mathcal{M}(\vec{q})}{4 m_\psi^2} \; .
\end{equation}
For the model discussed in this work, both the contribution from the (CP-even) scalar coupling as well as from the (CP-odd) pseudoscalar coupling can in principle contribute to the matrix element $\mathcal{M}(\vec{q})$. The former contribution induces a well-known Yukawa potential of the form
\begin{equation}
 V_S(r) =  \alpha_S \, e^{-m_\phi r}/r
\label{eq:VS}
\end{equation}
with $\alpha_S \equiv y_\psi^2 \cos^2 \delta_\psi / (4\pi)$. The pseudoscalar coupling, on the other hand, induces a Yukawa potential with a strongly suppressed coupling strength $\propto m_\phi^2 / m_\psi^2$, as well as further contributions to the potential scaling as $e^{-m_\phi r}/r^n$ with $n \geq 2$, which are of shorter range than the usual Yukawa potential induced by the exchange of a CP-even scalar. As discussed in detail in appendix~\ref{app:pseudoSIDM}, we therefore expect the scalar couplings to be the dominant source of DM self-interactions and use the potential from eq.~(\ref{eq:VS}) to calculate the momentum transfer cross section. This means in particular that we do not expect any sizeable effects from DM self-interactions in the case that the mediator has purely pseudoscalar couplings to DM.

A common approach to calculate $\sigma_\text{T}$ is to assume that the two interacting DM particles are distinguishable classical particles, which can interact with each other only via the $t$-channel exchange of a mediator~\cite{Tulin:2013teo,Agrawal:2016quu}. Under this assumption there is no need to differentiate between particle-particle and particle-antiparticle scattering and the differential cross section is independent of the total spin of the initial state. If the mediator is light and very weakly coupled, the differential cross section is then strongly peaked towards $\theta \rightarrow 0$. It is then a good approximation to replace the factor $1 - |\cos \theta|$ in eq.~(\ref{eq:sigT}) by $1 - \cos \theta$, which significantly simplifies the integration. However, this approximation is no longer valid if the mediator is not very light, if resonances become important or if the interference between $t$-channel and $u$-channel mediator exchange are relevant. Appendix~\ref{app:selfi} discusses how all of these effects can be consistently included and demonstrates that the differences to the standard approach are non-negligible.

In the presence of non-perturbative effects, the momentum transfer cross section depends on the relative DM velocity $v_\text{rel} \approx \sqrt{2} \, v_\text{disp}$ in such a way that larger effects can be expected for systems with small DM velocity dispersion $v_\text{disp}$ (such as dwarf galaxies) while constraints at larger velocities (e.g.\ from galaxy clusters) can be evaded. The strength of these constraints is a matter of ongoing debate (see e.g.~\cite{Kahlhoefer:2015vua,Wittman:2017gxn}), so we will adopt a rather conservative bound and require $\sigma_\mathrm{T} / m_\psi < 1 \, \mathrm{cm^2 \, g^{-1}}$ for $v = 1000\,\mathrm{km\,s^{-1}}$. On the scale of dwarf galaxies ($v \approx 30\,\mathrm{km \, s^{-1}}$), on the other hand, the observationally interesting range of self-interaction cross sections is $0.1\, \mathrm{cm^2 \, g^{-1}} \lesssim \sigma_\mathrm{T} / m_\psi  \lesssim 10 \, \mathrm{cm^2 \, g^{-1}}$~\cite{Zavala:2012us,Vogelsberger:2012ku,Kaplinghat:2013yxa,Vogelsberger:2014pda,Kaplinghat:2015aga}. We will compare this range to the various constraints that are discussed next.

\subsection{Indirect detection experiments and CMB constraints}

The first set of constraints stems from the same processes already discussed in section~\ref{sec:set-up}, namely the pair-annihilation of DM particles into mediators. Since these mediators are unstable, they will decay into SM final states, thus inducing potentially observable signatures. The Sommerfeld enhancement factor increases with decreasing DM velocity, so we can potentially expect strong constraints from indirect detection experiments probing $v/c \sim 10^{-4}\text{--}10^{-3}$. Even stronger constraints can result from CMB observations, which are sensitive to the DM annihilation rate at redshift $z \approx 1100$, where \mbox{$v/c \lesssim 10^{-7}$}~\cite{Bergstrom:2008ag,Mardon:2009rc, Galli:2009zc,Slatyer:2009yq,Zavala:2009mi,Feng:2010zp,Hannestad:2010zt,Finkbeiner:2010sm,Bringmann:2016din}.

Since we focus on the case $m_\phi < m_\psi$, only decays into SM final states are allowed and the branching ratios are independent of the magnitude of the couplings $y_\text{SM}$ and $y_\psi$. The ratio of the different decay modes does however depend on the mass of the mediator. For $m_\phi < 2 m_e$, only loop-induced decays into photons are kinematically accessible, while for $2 m_e < m_\phi < 2 m_\mu$ also decays into electrons play a relevant role. For larger mediator masses, both leptonic and hadronic decay modes can be relevant and uncertainties in the theoretical predictions are rather large. We will return to the question which branching ratios to use below.

In our scenario, the strongest constraints on present-day annihilation of DM particles are obtained from the very precise measurements of the positron flux made by the AMS-02 experiment \cite{PhysRevLett.113.121101,PhysRevLett.113.121102}. We adopt the bounds derived in \cite{Elor:2015bho} for one-step cascade annihilations, including the $4e$ and $4\mu$ final states. These bounds are independent of the mediator mass as long as $m_\phi \ll m_\chi$ and can be applied down to DM masses of 10 GeV.

To calculate CMB constraints, we assume that the Sommerfeld enhancement during recombination is fully saturated, which is a good approximation as long as the DM velocity during recombination satisfies $v_\text{rec} \lesssim m_\phi / (2 m_\psi)$~\cite{Bringmann:2016din}. To calculate the effect on the CMB, we multiply the resulting annihilation cross section with the appropriate efficiency factors from~\cite{Slatyer:2015jla}. The results are then compared with the most recent upper bound from Planck~\cite{Ade:2015xua}.

\subsection{Bounds on the Standard Model coupling}
\label{sec:SMbounds}

It is clear that a new mediating particle with mass below the GeV scale is only phenomenologically viable if it couples very weakly to the SM. A review of the most relevant constraints can be found in~\cite{Krnjaic:2015mbs} for the case of scalar couplings and in~\cite{Dolan:2014ska} for the case of pseudoscalar couplings (see also~\cite{Alekhin:2015byh} for an experimental proposal to improve these constraints in the near future). The strongest constraints for mediator masses $m_\phi < 100\,\mathrm{MeV}$ typically come from searches for rare kaon decays. The reason is that $W$-boson loops induce effective flavour-changing interactions of the form
\begin{equation}
 \mathcal{L}_\text{FCNC} \supset h^S_{ds} \, \phi \, \bar{d} s + \text{h.c.} \; ,
\end{equation}
where for $\delta_\text{SM} = 0$~\cite{Batell:2009jf}
\begin{equation}
h^S_{ds} \approx \frac{3 \alpha \, y_\text{SM} \, m_s \, m_t^2}{32 \pi \, m_W^2 \, \sin(\theta_W)^2 \, \vev} \, V_{ts} V_{td}^\ast
\end{equation}
with $m_W$ the $W$-boson mass, $\alpha \equiv e^2 / (4 \pi)$, $\theta_W$ the Weinberg angle and $V$ the CKM matrix. For $\delta_\text{SM} \neq 0$, the $W$-boson loop is divergent, indicating a sensitivity to the specific UV completion~\cite{Freytsis:2009ct}. To estimate the magnitude of the expected effects, one can introduce a cut-off at a specific new-physics scale $\Lambda$. In this case, one obtains for $\delta_\text{SM} = \pi/2$~\cite{Dolan:2014ska}
\begin{equation}
h^S_{ds} \approx - \frac{\alpha \, y_\text{SM} \, m_s \, m_t^2}{8 \pi \, m_W^2 \, \sin(\theta_W)^2 \, \vev} \, V_{ts} V_{td}^\ast \, \log\left(\frac{\Lambda^2}{m_t^2}\right) \; . \end{equation}

These new interactions can induce new kaon decay modes~\cite{Deshpande:2005mb}:
\begin{equation}
\Gamma(K^+ \rightarrow \pi^+ \phi) = \frac{1}{16 \pi \, m_{K^+}^3} \lambda^{1/2}(m_{K^+}^2, m_{\pi^+}^2, m_\phi^2) \left(\frac{m_{K^+}^2 - m_{\pi^+}^2}{m_s - m_d}\right)^2 |h^S_{ds}|^2 \; ,
\label{eq:Kwidth}
\end{equation}
where $\lambda(a,b,c) =(a-b-c)^2-4\,b\,c$ and we neglect a form factor $|f_0^{K^+}(m_\phi^2)|^2$, which is close to unity~\cite{Marciano:1996wy}. Since the light mediator is rather long-lived, it will typically escape from the detector without decaying. Dedicated searches for this decay mode by experiments like E787~\cite{Adler:2004hp} and E949~\cite{Anisimovsky:2004hr,Artamonov:2009sz} place an upper bound of 
\begin{equation}
\text{BR}(K^+ \rightarrow \pi^+ + \phi(\to\text{inv})) \lesssim 5 \cdot 10^{-11} \; .
\label{eq:invwidth}
\end{equation}
This bound implies
\begin{align}
y_\text{SM} \lesssim 1.9 \cdot 10^{-4} & \qquad \text{for } \delta_\text{SM} = 0 \; ,\\
y_\text{SM} \log\left(\Lambda^2/m_t^2\right) \lesssim 1.6 \cdot 10^{-4} & \qquad \text{for } \delta_\text{SM} = \pi /2 \; 
\end{align} 
for $m_\phi \lesssim 50\,\mathrm{MeV}$. We note that the NA62 experiment~\cite{Ceccucci:2014oza} is expected to measure the branching ratio for $K^+ \to \pi^+ \bar{\nu} \nu$ with a precision of 10\% and thereby significantly improve the bounds on new invisible decay modes.

At the same time, astrophysical and cosmological constraints can be used to place a lower bound on the coupling $y_\text{SM}$. Most importantly, if the mediators produced in the early Universe have a lifetime $\tau_\phi \gtrsim 1\,\mathrm{s}$, they will decay during or after BBN and can thereby potentially spoil the successful prediction of the abundance of the various elements. While it is challenging to accurately calculate the magnitude of these effects, it is clear that BBN constraints can be evaded if the mediator decays sufficiently quickly. As long as decays into leptons are kinematically allowed, a sufficiently short lifetime can always be achieved without conflicting with the upper bounds on $y_\text{SM}$ mentioned before. For purely photonic decays, however, it is typically impossible for phenomenologically viable mediators to decay before BBN. A dedicated study of BBN constraints is imperative to determine whether such mediators can be consistent with the observed abundance of elements.

Strong constraints for sub-MeV mediators also come from astrophysical observations, such as the lifetime of horizontal branch stars~\cite{Raffelt:1987yu} and the duration of the neutrino signal from SN1987a (see~\cite{Krnjaic:2015mbs} for a recent re-analysis). To be safe from these constraints, we limit ourselves to $m_\phi > 30\,\mathrm{keV}$ and $y_\text{SM} > 3 \cdot 10^{-5}$ unless explicitly stated otherwise.

\subsection{Direct detection experiments and thermalisation}
\label{sec:DD}

The scenario considered in this work is very strongly constrained by direct detection experiments. The reason is that the Lagrangian given in eqs.~(\ref{eq:LDM}) and~(\ref{eq:Lmix}) leads to spin-independent DM-nucleon scattering, which is significantly enhanced by the small mass of the mediator $\phi$. The corresponding scattering cross section of dark matter off a target nucleus with mass $m_T$ and mass number $A$ is given by
\begin{equation}
 \frac{\text{d}\sigma^\text{SI}_T}{\text{d}E_R} = \frac{f_p^2 \, m_p^2}{2 \pi \, \vev^2} \frac{m_T A^2 F^2(E_R)}{v^2} \frac{y_\psi^2 \, y_\text{SM}^2 \, \cos^2 \delta_\psi \, \cos^2 \delta_{\text{SM}}}{(m_\phi^2 + q^2)^2} \; ,
\label{eq:sigN}
\end{equation}
where $f_p = f_n \approx 0.3$ is the effective nucleon coupling~\cite{Cline:2013gha}, $v$ is the DM velocity, $F^2(E_R)$ is the usual form factor for spin-independent scattering and $q = \sqrt{2 m_T E_R} \sim (1\text{--}100)\,\mathrm{MeV}$ denotes the momentum transfer in a nuclear recoil event. For $m_\phi \lesssim q$, the scattering process can effectively be treated as a long-range interaction.

The strongest constraints on $\sigma_N$ come from LUX~\cite{Akerib:2015rjg,Akerib:2016vxi} and PandaX~\cite{Tan:2016zwf} for DM masses above 5 GeV and from CRESST-II~\cite{Angloher:2015ewa,Angloher:2017zkf} and CDMSlite~\cite{Agnese:2015nto} for smaller masses. We re-analyse the 2015 results from LUX~\cite{Akerib:2015rjg} following~\cite{Catena:2016hoj} to take into account the modified shape of the nuclear recoil spectrum due to the transition from contact interactions to long-range interactions. For CRESST-II, we consider the first bin ($0.3\,\mathrm{keV} \leq E_R \leq 0.4\,\mathrm{keV}$) of the data presented in~\cite{Angloher:2015ewa}, and compare the six observed events to the number of expected events calculated from the efficiencies provided in~\cite{Angloher:2017zkf}. We conservatively only include recoil energies above $0.114$ keV in our calculation, corresponding to three standard deviations in the energy resolution, and calculate the upper bound by assuming no background events.

The expression for the DM-nucleon scattering cross section in eq.~(\ref{eq:sigN}) suggests that constraints from direct detection experiments become arbitrarily weak as $\delta_\text{SM} \to \pi/2$ or $\delta_\psi \to \pi/2$. While it is true that standard spin-independent interactions vanish in this limit, a non-zero contribution arises from interaction terms that are suppressed by additional powers of $q^2/m_N^2$ or $q^2 / m_\psi^2$~\cite{Fan:2010gt}. To include these effects, we express the scattering amplitude in terms of a complete set of non-relativistic operators~\cite{Fitzpatrick:2012ix} and calculate experimental constraints taking into account the full momentum dependence. We find that for $\delta_\text{SM} = \pi / 2$ direct detection experiments do not yield any relevant constraints on the model parameter space irrespective of the value of $\delta_\psi$.

For $\delta_\text{SM} = 0$, on the other hand, the constraints from direct detection experiments are so strong that they put into question one of our fundamental assumptions, namely that the dark sector was in thermal equilibrium with the SM sector at high temperatures. For this to happen, we must require that the DM production rate at some point in the early Universe exceeded the Hubble expansion rate:
\begin{equation}
 n_f(T) \langle \sigma(f \bar{f} \to \psi \bar{\psi}) v \rangle \gtrsim H(T) \; ,
\end{equation}
where $n_f(T)$ is the fermion number density as a function of the temperature $T$ and the brackets denote thermal averaging~\cite{Gondolo:1990dk}. The Hubble rate is given by $H(T) \simeq 1.66 \sqrt{g_\ast} T^2 / m_\text{Pl}$ in terms of the effective number of degrees of freedom $g_\ast$ and the Planck mass $m_\text{Pl}$. This requirement, which is most easily satisfied for $T \approx \max(m_t, m_\psi)$, can be used to obtain a lower bound on the product of the two couplings:
\begin{equation}
y_\text{SM} \, y_\psi \gtrsim \begin{cases} 1.1 \cdot 10^{-6} &\mbox{for } m_\psi \lesssim m_t \\
\left(\dfrac{m_\psi}{\text{GeV}}\right)^{1/2} \cdot \dfrac{9.4 \cdot 10^{-8}}{\sqrt{1.9 - 0.26 \, (1 + \cos 2 \delta_\psi)}} & \mbox{for }  m_\psi \gtrsim m_t \end{cases} \quad (\delta_\text{SM} = 0)  \; .
\end{equation}
Whenever this lower bound is in conflict with the upper bound obtained from direct detection experiments, the conclusion is that a different mechanism must be responsible for bringing the two sectors into thermal equilibrium in the early Universe.\footnote{For example, in the toy model introduced in appendix~\ref{app:toymodel}, thermalisation can also occur via the heavy pseudoscalar $A$, in which case DM production can be resonantly enhanced.} We will return to this issue in the next section.

\section{Results}
\label{sec:results}

Having discussed the various relevant observables and constraints, we now present the results of our analysis and show the viable regions of parameter space. We are particularly interested in finding parameter regions where large self-interactions on the scale of dwarf galaxies are consistent with all exclusion bounds. We will begin by revisiting the frequently studied case of purely scalar interactions, i.e.\ $\delta_\text{SM} = \delta_\psi = 0$. We show that in this case direct detection constraints are so strong that it is almost impossible to simultaneously satisfy all phenomenological requirements. A possible way to avoid these constraints would be to consider purely pseudoscalar interactions ($\delta_\text{SM} = \delta_\psi = \pi/2$), but in this case one does not obtain large DM self-interactions (see section~\ref{sec:SIDM}). Instead, we will therefore consider the case where $\delta_\text{SM} \approx \pi / 2$ but $\delta_\psi$ is arbitrary, so that direct detection constraints are suppressed but sizeable self-interaction cross sections can be obtained.

\subsection{Purely scalar interactions}

\begin{figure}[t]
\vspace*{-1.6cm}
\hspace*{-0.95cm}
\includegraphics[scale=0.95]{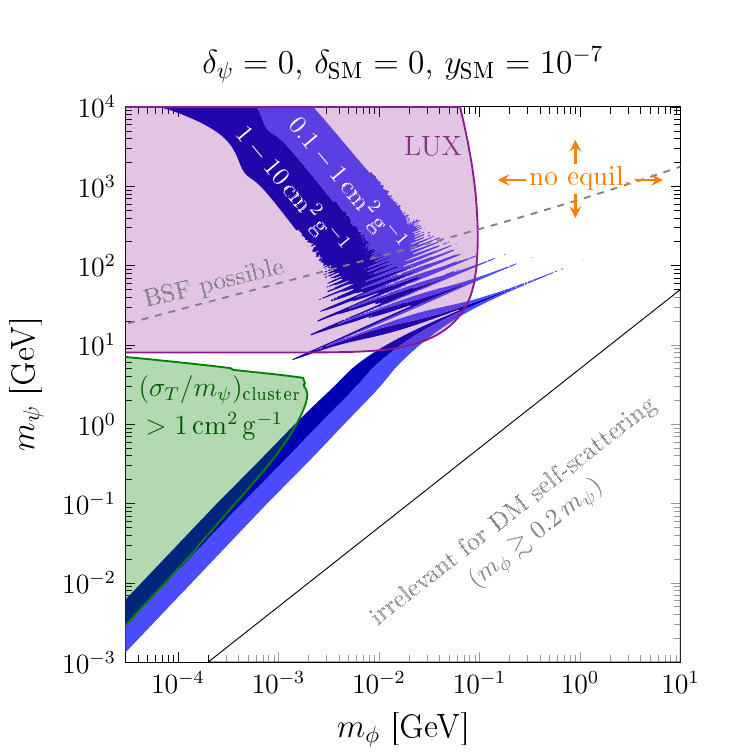}\hspace{1.2cm}
\includegraphics[scale=0.95]{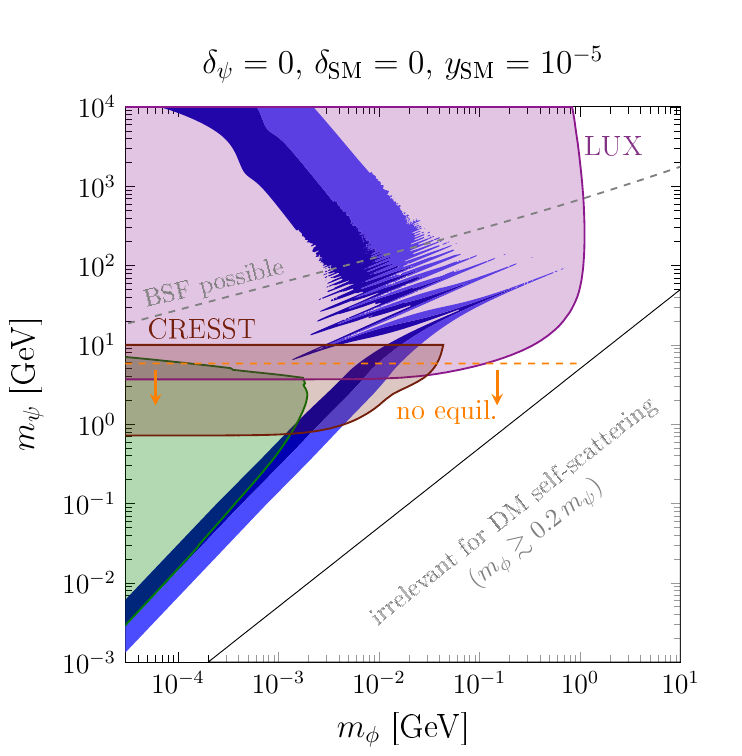}\\[0.05cm]
\hspace*{-0.9cm}
\includegraphics[scale=0.95]{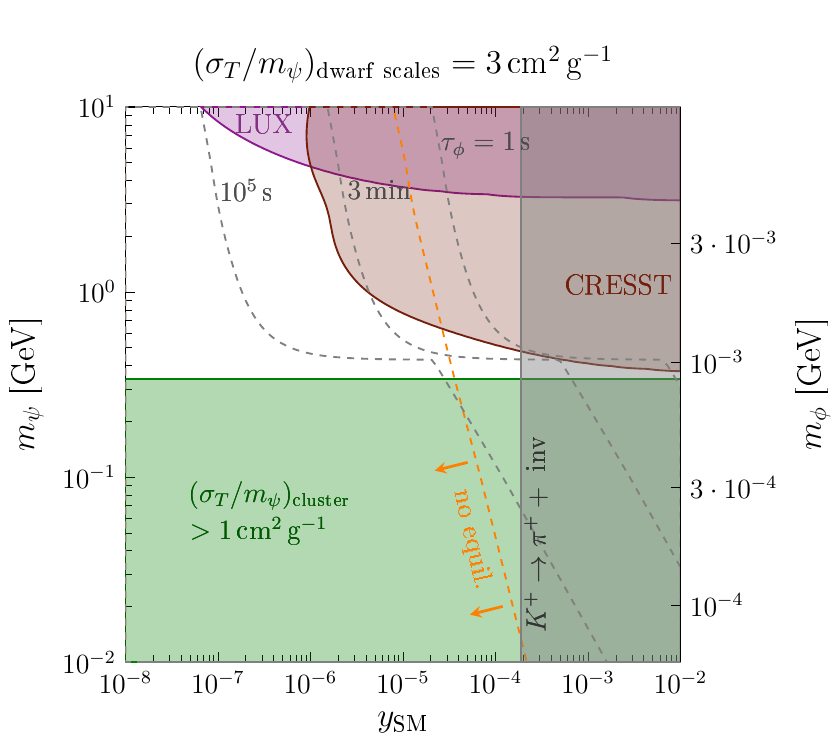}\hspace{0.3cm}
\includegraphics[scale=0.95]{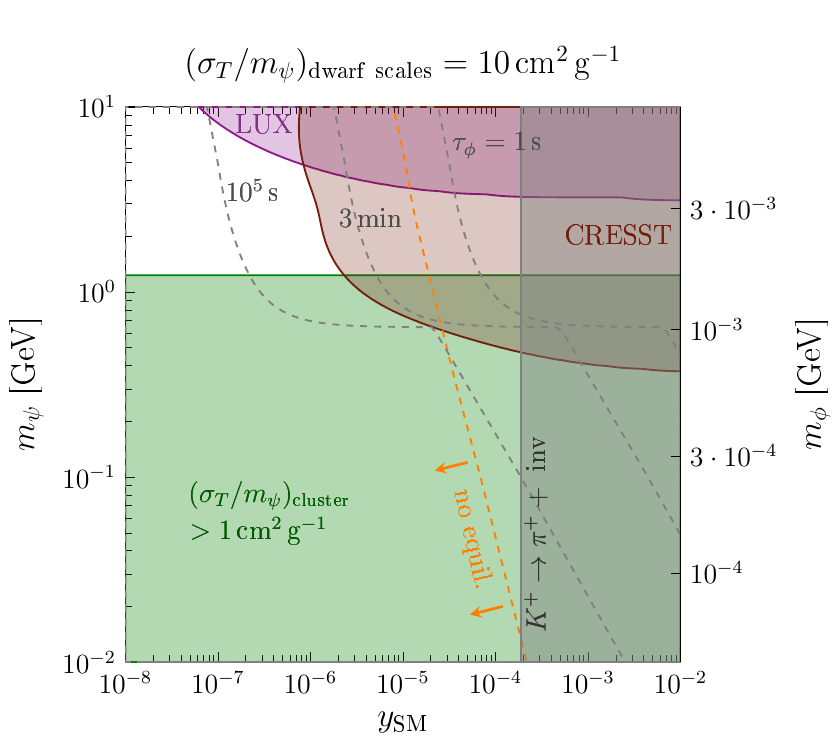}\\[0.05cm]
\caption{Constraints and interesting parameter regions for the case of purely scalar couplings both to DM and to SM fermions ($\delta_{\psi} = \delta_{\text{SM}} = 0$) for fixed SM coupling $y_\text{SM}$ (top row) and fixed self-interaction cross section $\sigma_\mathrm{T} / m_\psi$ (bottom row). In all panels $y_{\psi}$ is fixed by the relic density requirement. Note that for $10^{-7} \lesssim y_\text{SM} \lesssim 10^{-4}$ and $m_\phi \lesssim 0.1 \,\mathrm{GeV}$ constraints from SN1987a (not shown) may also become relevant.}
\label{fig:pure_scalar}
\end{figure}

For purely scalar interactions direct DM annihilation proceeds exclusively via $p$-wave, and hence does not lead to relevant constraints from indirect detection experiments even when taking into account Sommerfeld enhancement. It has been noted that bound-state formation (BSF) can still potentially proceed via $s$-wave, leading to additional constraints not included in our calculation~\cite{An:2016kie}. These can potentially be important for specific combinations of $m_\psi$ and $m_\phi$ in regions of the parameter space where $\alpha_S^2 m_\psi/m_\phi \gtrsim 4$, with $\alpha_S \equiv y_\psi^2 \cos^2 (\delta_\psi)/4 \pi$~\cite{An:2016kie,Cirelli:2016rnw}. This is indicated by the gray dashed lines in Figs.~\ref{fig:pure_scalar}-\ref{fig:mixed_largedelta}.


The crucial question therefore is whether the bounds from direct detection experiments are compatible with the requirement of large DM self-interactions as well as with the lower bounds on the couplings from the mediator lifetime and the thermalisation condition for the dark sector. We show the preferred and excluded parameter regions for the case of purely scalar interactions in figure~\ref{fig:pure_scalar}. At each point, we have fixed $y_\psi$ by the requirement to reproduce the observed relic abundance as described in section~\ref{sec:relic}. In the top row, we consider fixed values of the SM coupling $y_\text{SM}$ and vary the two masses $m_\psi$ and $m_\phi$. Constraints from direct detection are shown in purple and brown, whereas the constraint $\sigma_\mathrm{T} / m_\psi < 1 \, \mathrm{cm^2 / g}$ on cluster scales is shown in green. The parameter regions corresponding to large self-interaction cross sections on dwarf galaxy scales are shown in blue: The darker region corresponds to $1 \, \mathrm{cm^2 g^{-1}} < \sigma_\mathrm{T} / m_\psi < 10 \, \mathrm{cm^2 g^{-1}}$, the lighter to $0.1 \, \mathrm{cm^2 g^{-1}} < \sigma_\mathrm{T} / m_\psi < 1 \, \mathrm{cm^2 g^{-1}}$. In addition, we show in orange the region of the parameter space where the mediator is too weakly coupled to bring the two sectors into thermal equilibrium.

For DM masses larger than about 5 GeV, we find direct detection bounds to be so strong that it is essentially impossible to obtain large self-interaction cross sections unless $y_\text{SM}$ is extremely small, in which case a different mechanism must be responsible for bringing the dark sector and the visible sector into thermal equilibrium and late decays of the mediator are unavoidable. For DM masses below about 0.5 GeV, on the other hand, we find that the DM self-interaction cross section is essentially independent of the DM velocity. This parameter region is phenomenologically less interesting, as it is impossible to obtain larger effects on dwarf galaxy scales than on galaxy cluster scales.

The most interesting parameter region therefore corresponds to DM masses of around 1 GeV, which is precisely the parameter region probed by novel direct detection experiments with very low threshold, such as CRESST-II. We zoom into this parameter region in the bottom row of figure~\ref{fig:pure_scalar} and, rather than fixing $y_\text{SM}$, impose a fixed value of $\sigma_\mathrm{T} / m_\psi$ on dwarf galaxy scales. This approach effectively imposes a relation between $m_\psi$ and $m_\phi$, as indicated by the two different y-axes.\footnote{Note that this approach only works for $m_\psi \lesssim 10\,\text{GeV}$, because for larger DM masses resonances become important and it is no longer possible to fix $m_\phi$ uniquely as a function of $m_\psi$. Moreover, for $m_\psi \lesssim 10\,\text{GeV}$ the velocity dependence of $\sigma_\mathrm{T} / m_\psi$ saturates well above the scale relevant for dwarf galaxies and therefore the plots shown in the bottom row of figure~\ref{fig:pure_scalar} do not depend on the precise value assumed for $v_\text{rel}$.} These plots clearly demonstrate that it is impossible to obtain $\sigma_\mathrm{T} / m_\psi = 10 \, \mathrm{cm^2 g^{-1}}$ on dwarf galaxy scales for any combination of couplings and masses consistent with all other requirements. Self-interaction cross sections of the order of $3 \, \mathrm{cm^2 g^{-1}}$ are possible in a small region of parameter space around $m_\psi \approx 0.5\,\mathrm{GeV}$, $m_\phi \approx 1\,\mathrm{MeV}$ and $y_\text{SM} \approx 10^{-4}$. Note however that in this finely tuned parameter region the mediator has a life time $\gtrsim 1$ s, requiring a dedicated study of constraints from BBN to determine whether this corner of parameter space is still viable. Intriguingly, this window may also soon be probed by measurements of rare kaon decays at NA62~\cite{Ceccucci:2014oza}.

\subsection{Mixed (pseudo)scalar interactions}

We have shown above that direct detection constraints are now essentially so strong that it is impossible for the purely scalar case to obtain self-interaction cross sections as large as $10 \, \mathrm{cm^2 g^{-1}}$ on dwarf galaxy scales. We therefore now turn to the case where $\delta_\text{SM} \approx \pi/2$, i.e.\ the interaction of $\phi$ with SM fermions are nearly CP-conserving, while $\delta_\psi$ is allowed to take arbitrary values. As discussed in appendix~\ref{app:toymodel}, this set-up can be obtained naturally from spontaneous CP violation as a consequence of the weak coupling between the dark and the visible sector. Direct detection constraints then do not impose any relevant constraints on the remaining parameters (see section~\ref{sec:DD}).\footnote{We have checked that this statement remains true even if we set $\delta_\text{SM} = \pi/2 - y_\text{SM} / y_\psi$, corresponding to the amount of CP violation expected in the model discussed in appendix~\ref{app:toymodel}.}

Allowing arbitrary values for $\delta_\psi$ means that indirect detection constraints become important. The origin of these constraints can be immediately understood from eq.~(\ref{eq:2to2}) and figure~\ref{fig:relic_density}. As the $s$-wave annihilation cross section scales as $\sin^2(2 \delta_\psi)$, for $\delta_\psi \approx 0$ or $\delta_\psi \approx \pi/2$ the relic density is dominantly set via $p$-wave processes, which are irrelevant during recombination and for present-day DM searches. The $s$-wave contribution, on the other hand, is negligibly small (or in fact zero in the case of purely scalar or pseudoscalar interactions) both during thermal freeze-out and at later times, even when including Sommerfeld enhancement. As soon as we move significantly away from the two limiting cases, however, the relic density is dominantly set via $s$-wave processes and one obtains relevant constraints from CMB measurements and indirect detection experiments. To calculate these constraints in detail, we take the decay widths and branching ratios for a spin-0 mediator with pseudoscalar couplings to SM fermions from~\cite{Dolan:2014ska}. As shown for the case of a vector mediator in~\cite{Bringmann:2016din} these constraints are typically so strong that they completely exclude the possibility to obtain large DM self-interactions.

\begin{figure}[t]
\hspace*{0.1cm}
\includegraphics[scale=0.95]{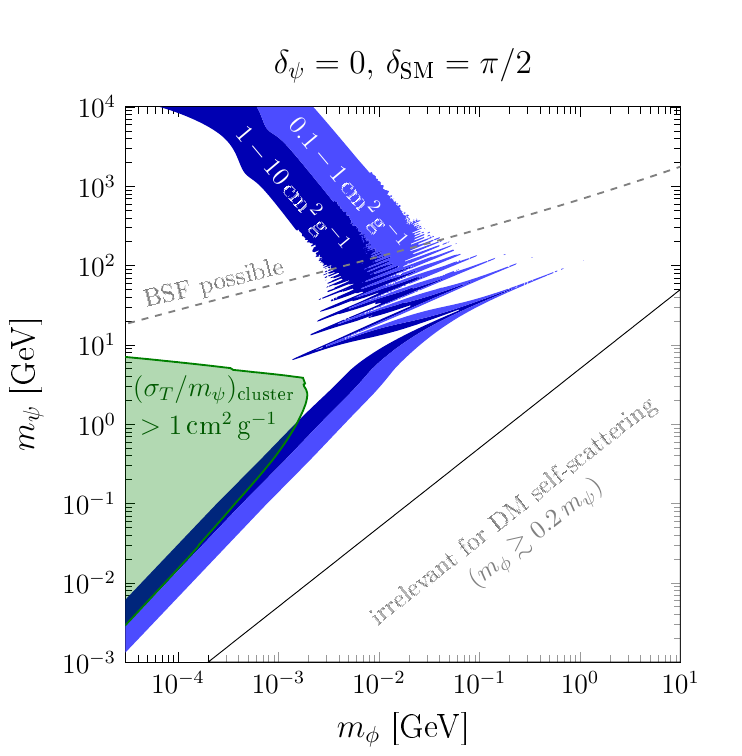}\hspace{0.7cm}
\includegraphics[scale=0.95]{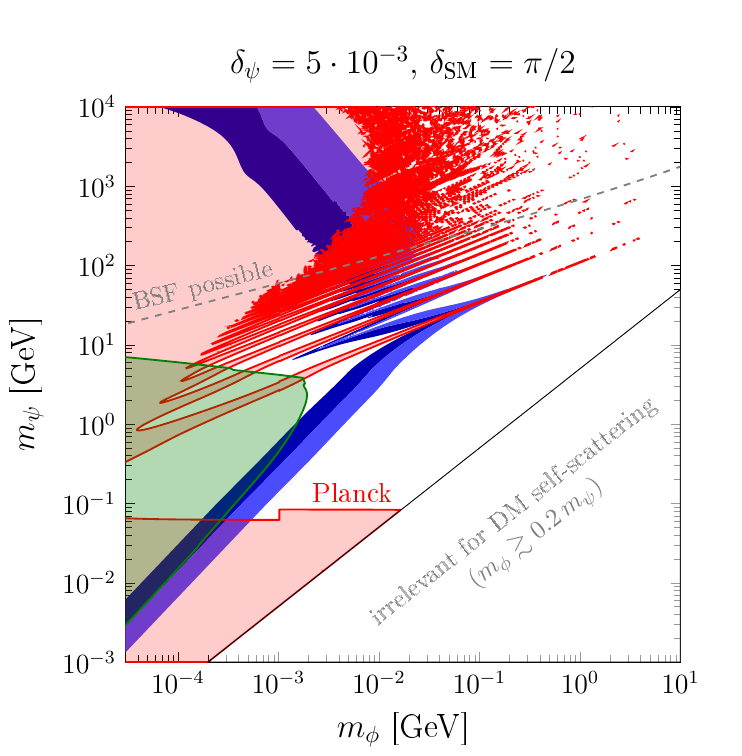}\\[0.5cm]
\hspace*{0.1cm}
\includegraphics[scale=0.95]{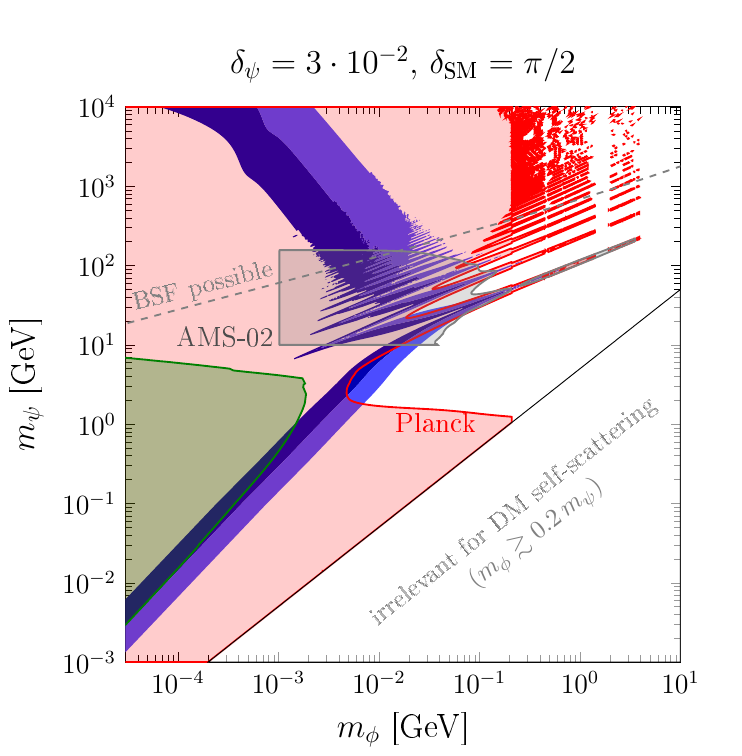}\hspace{0.7cm}
\includegraphics[scale=0.95]{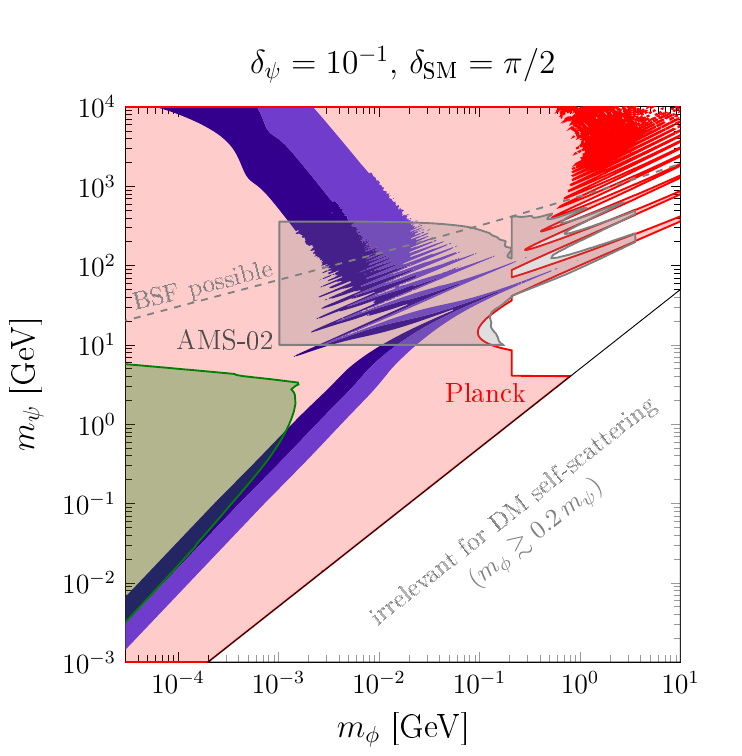}
\caption{Constraints and interesting parameter regions for the case $\delta_\text{SM} = \pi / 2$ and $\delta_\psi \approx 0$. In all panels $y_{\psi}$ is fixed by the relic density requirement.}
\label{fig:mixed_smalldelta}
\end{figure}

\begin{figure}[t]
\hspace*{0.1cm}
\includegraphics[scale=0.95]{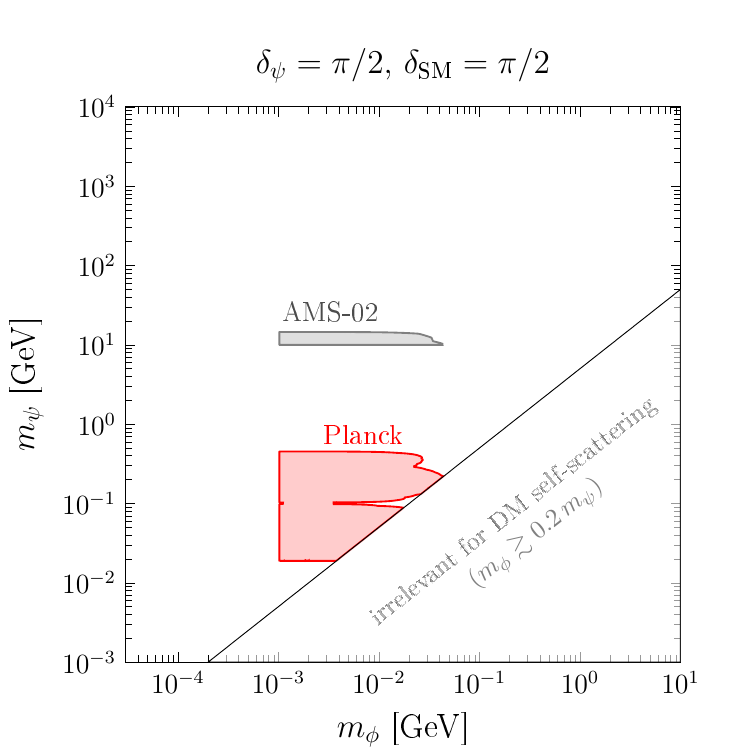}\hspace{0.7cm}
\includegraphics[scale=0.95]{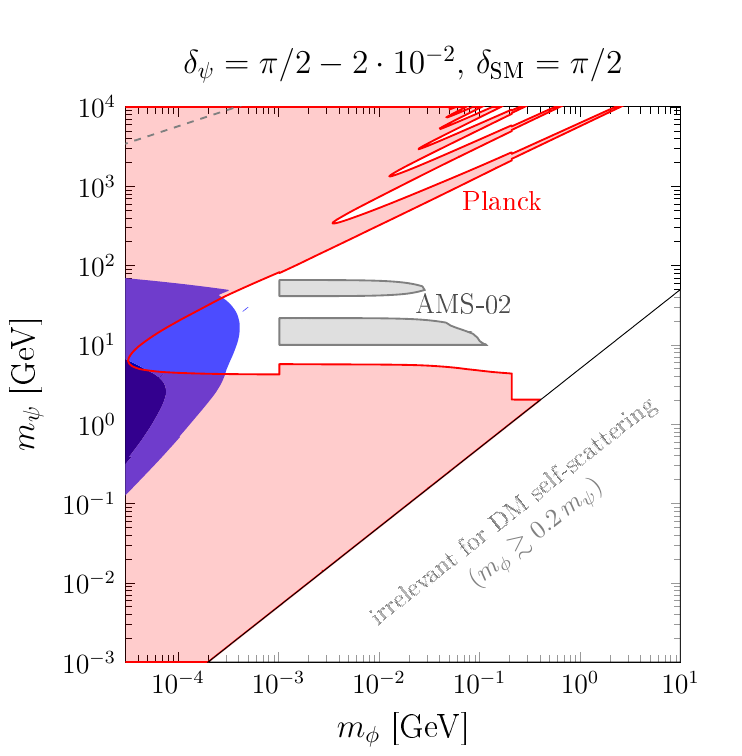}\\[0.5cm]
\hspace*{0.1cm}
\includegraphics[scale=0.95]{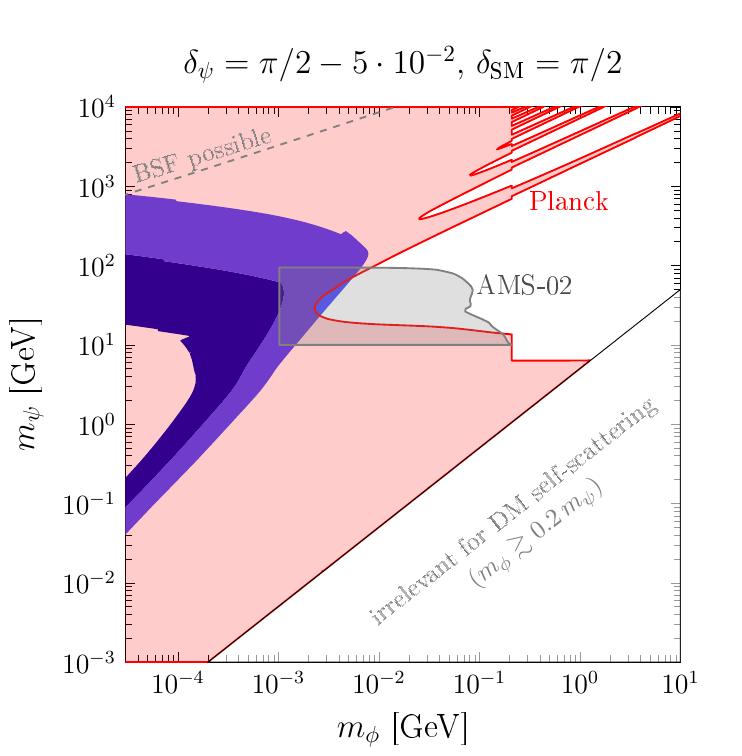}\hspace{0.7cm}
\includegraphics[scale=0.95]{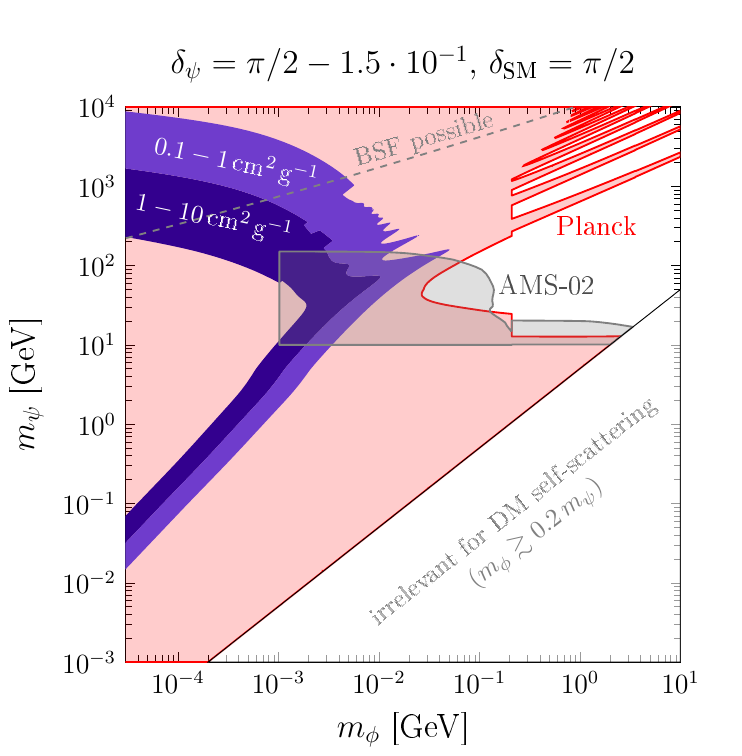}
\caption{Same as figure~\ref{fig:mixed_smalldelta} but for $\delta_\psi \approx \pi/2$.}
\label{fig:mixed_largedelta}
\end{figure}

The most interesting parameter regions are therefore the ones corresponding to either $\delta_\psi \approx 0$ or $\delta_\psi \approx \pi/2$. The first case is shown in figure~\ref{fig:mixed_smalldelta} for four different values of $\delta_\psi \in \left[ 0, 0.1\right]$, while the latter case is shown in figure~\ref{fig:mixed_largedelta} for four different values of $\delta_\psi \in \left[ \pi/2-0.15, \pi/2\right]$. Since direct detection constraints are irrelevant in these figures, it is not necessary to specify $y_\text{SM}$. In particular, it is always possible for this coupling to be sufficiently large to ensure thermalisation of the two sectors and to avoid late decays of the mediator provided $m_\phi > 2 m_e$.

The crucial difference between these two cases is that the Sommerfeld enhancement factor depends only on the scalar component of the DM-mediator coupling, $y_\psi \cos \delta_\psi$, and is therefore much larger for $\delta_\psi \approx 0$ than for $\delta_\psi \approx \pi/2$. We therefore expect much larger self-interaction cross sections~-- and much stronger constraints from indirect detection experiments~-- in the former case than in the latter. Indeed, the parameter regions corresponding to large self-interactions shown in figure~\ref{fig:mixed_smalldelta} are very similar to the ones previously shown in figure~\ref{fig:pure_scalar}. In the case $\delta_\psi = 0$, shown in the top-left panel, there are no relevant indirect detection constraints, so large DM self-interactions are phenomenologically viable. However, since we consider $\delta_\text{SM} = \pi/2$ we cannot invoke CP symmetry to ensure that $\delta_\psi = 0$ holds exactly. It is therefore crucial to understand how large $\delta_\psi$ can be before the scenario is ruled out by indirect detection constraints. Indeed, for $\delta_\psi = 0.1$ the entire parameter region corresponding to large self-interactions is in conflict with CMB constraints (bottom-right panel). In order to find allowed parameter space we must require $\delta_\psi$ to be of order $10^{-2}$ or smaller (see top-right and bottom-left panel).

For $\delta_\psi \approx \pi/2$ we encounter a different situation, because large self-interactions are only possible if $\delta_\psi$ is appreciably different from the purely pseudoscalar case. In particular, the resonant enhancement of self-interactions found in the case $\delta_\psi \approx 0$ is absent unless $\delta_\psi \lesssim \pi/2 - 0.1$. For such values of $\delta_\psi$, however, there are significant constraints from CMB measurements (see bottom-right panel of figure~\ref{fig:mixed_largedelta}). Indeed, in none of the four panels shown in figure~\ref{fig:mixed_largedelta} is it possible to obtain DM self-interaction cross sections significantly larger than $1 \, \mathrm{cm^2 g^{-1}}$ without being excluded by CMB constraints. Note, however, that for $\delta_\psi \approx \pi/2$ the velocity dependence of the self-interactions has typically not saturated at dwarf galaxy scales, so that the precise value of the self-scattering cross section depends on the assumed DM relative velocity.

\begin{figure}
\hspace*{-0.6cm}
\includegraphics[scale=0.95]{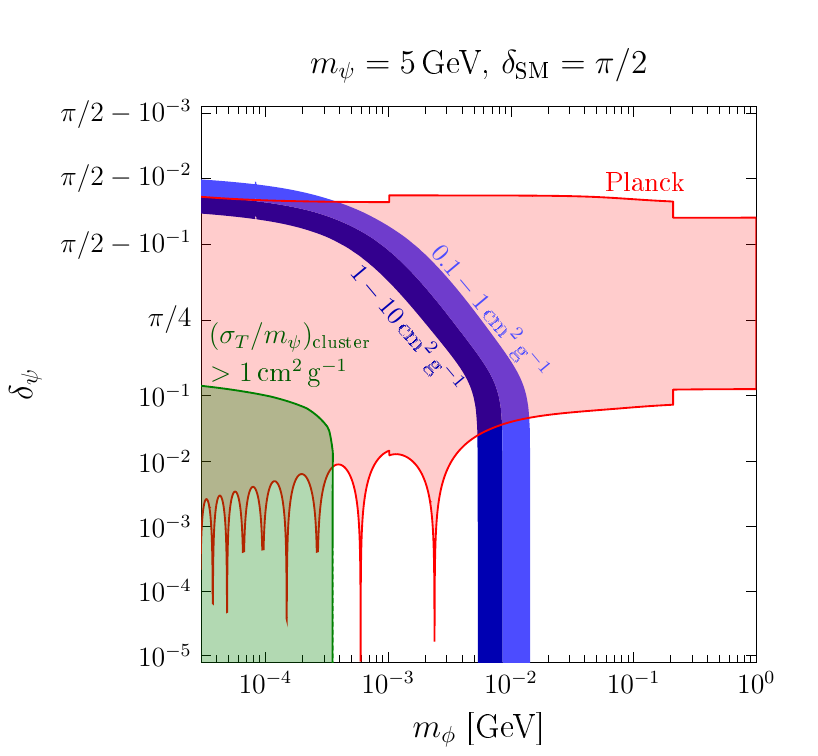}\hspace{0.06cm}
\includegraphics[scale=0.95]{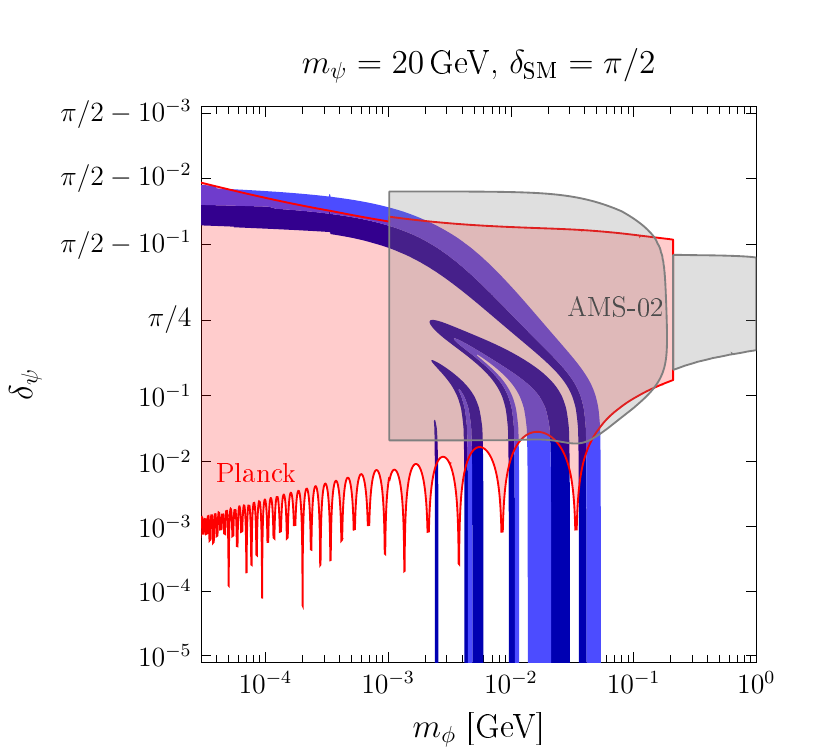}
\caption{Same as figure~\ref{fig:mixed_smalldelta} and~\ref{fig:mixed_largedelta} but showing the constraints as a function of $\delta_\psi$ for fixed $m_\psi$. Note the special scaling of the y-axis intended to emphasize the interesting limiting cases $\delta_\psi = 0$ and $\delta_\psi = \pi/2$.}
\label{fig:mixed_deltapsi_vs_mphi}
\end{figure}

To understand whether our conclusions depend on the specific choices of $\delta_\psi$ made in figure~\ref{fig:mixed_smalldelta} and~\ref{fig:mixed_largedelta}, we can show the constraints as a function of $\delta_\psi$ for fixed DM masses. This is done in figure~\ref{fig:mixed_deltapsi_vs_mphi} for $m_\psi = 5\,\text{GeV}$ and $m_\psi = 20\,\text{GeV}$, varying $\delta_\psi$ in the complete range between $0$ and $\pi/2$. As expected, indirect detection constraints are overwhelming in the range $0.1 \lesssim \delta_\psi \lesssim \pi/2 - 0.1$. For larger $\delta_\psi$, self-interaction cross sections as large as $\gtrsim 1\,\text{cm}^2/\text{g}$ are marginally compatible with CMB measurements. Values of $\delta_\psi$ smaller than 0.1, on the other hand, make it possible to reconcile the requirement of large self-interactions on dwarf galaxy scales with all indirect detection constraints.

To conclude this section, let us briefly discuss how our results would change if we were to allow arbitrary values of $\delta_\text{SM}$. First of all, varying $\delta_\text{SM}$ would affect the branching ratios of the mediator for $m_\phi > 2 m_\pi$, as hadronic decay modes are more important for scalar than for pseudoscalar couplings. The resulting changes in the CMB constraints do however not modify any of our central conclusions. The more important difference clearly comes from the fact that direct detection constraints become relevant again. For example, for $m_\psi = 5\,\text{GeV}$, $m_\phi = 10\,\mathrm{MeV}$, $\delta_\psi = 10^{-2}$ and $y_\text{SM} = 10^{-4}$, direct detection constraints will become relevant as soon as $\delta_\text{SM} \lesssim \pi/2 - 6 \cdot 10^{-3}$. In other words, for this solution to be viable a mechanism like the one discussed in appendix~\ref{app:toymodel} must ensure that $\delta_\text{SM}$ is indeed very close to $\pi / 2$.

\section{Conclusions}
\label{sec:conclusions}

In this work we have discussed the detailed phenomenology of DM
particles interacting with each other via the exchange of a very light
spin-0 mediator. Such a set-up is motivated by the observation that the
resulting DM self-interactions can be sufficiently large to affect
astrophysical systems and at the same time exhibit a characteristic
velocity dependence which makes it possible to consistently describe
astrophysical observations over a range of different scales.
Furthermore, thermal freeze-out can be naturally realised in models with
a light mediator and the observed DM relic abundance can be obtained
via direct annihilation of a DM pair into light mediators.

In a minimal realisation of the set-up, the mediator must also couple to
SM states in order to bring the dark sector into thermal equilibrium
with the SM sector and to ensure that the mediator decays before
BBN. In this case, one can expect a number of
relevant constraints from a variety of experimental and observational
probes. Specifically, we have considered direct and indirect detection
experiments, as well as CMB measurements and bounds from searches for
rare meson decays, and compared the resulting constraints to the
parameter regions relevant for DM self-interactions.

For the case of purely scalar interactions, we have pointed out the important role of
low-threshold direct detection experiments like CRESST-II. These
searches strongly constrain the viable regions of parameter space and
essentially make it impossible to obtain self-interaction cross sections
larger than $5\,\mathrm{cm^2 \, g^{-1}}$ on dwarf galaxy scales and
smaller than $1\,\mathrm{cm^2 \, g^{-1}}$ on galaxy cluster scales in
the simplest set-up. For purely pseudoscalar interactions on the other hand no relevant DM 
self-interactions are expected to arise.

Large self interactions with substantially suppressed constraints can be realised however  
if the mediator couples to DM and SM states with different CP phases
$\delta_\psi$ and $\delta_\text{SM}$, respectively. In particular, for
$\delta_\text{SM} \approx \pi/2$, direct detection constraints are
largely absent. In the absence of CP conservation, however, indirect
detection constraints and CMB constraints are very strong and impose
$\delta_\psi \lesssim 0.1$ or $\delta_\psi \gtrsim \pi/2 - 0.1$. In
particular in the former case it is possible to obtain large DM
self-interactions on dwarf galaxy scales consistent with all other requirements.

There are a number of ways in which the tension between the different
constraints and requirements discussed in this work can be ameliorated.
For example, for the case of purely scalar interactions constraints from
direct detection experiments may be reconciled with BBN constraints in
models where the mediator has suppressed couplings to nucleons but can
still decay sufficiently quickly into light leptons and photons (e.g.\
leptophilic DM~\cite{Fox:2008kb}). In such a model one would likely need
a different mechanism for bringing the dark sector into thermal
equilibrium with the SM, which offers the interesting possibility to
have different temperatures in the two sectors. In fact, it is also
possible to obtain SIDM from a dark sector that was never in thermal
equilibrium with the SM, for example via the freeze-in
mechanism~\cite{Bernal:2015ova}, so that direct and indirect detection
constraints are absent. Finally, it is worth pointing out that both BBN
and CMB constraints can be significantly weakened if DM is asymmetric~\cite{Baldes:2017gzw} or if the mediator decays
into inert particles such as sterile neutrinos~\cite{Aarssen:2012fx}.
Exploring cosmological constraints for such set-ups offer an
interesting avenue for future research.

In conclusion, while the idea of sizeable DM self-interactions remains
very attractive, the simplest attempts to construct specific models
based on light mediators face a number of strong constraints. Following
the recent observation that vector mediators are in strong tension with
CMB constraints and indirect detection
experiments~\cite{Bringmann:2016din}, we have extended previous 
analyses in this work to show that
also the case of scalar mediators is essentially ruled out in its
simplest realization. While extended models, such as mediators with
CP-violating couplings, are presently still viable, the exciting
interplay of astrophysical observables and particle physics experiments
means that we can hope to comprehensively explore the idea of DM
particles interacting via the exchange of light mediators in the near
future.

\acknowledgments

We thank Iason Baldes, Torsten Bringmann, Jim Cline, Michael Duerr, Norbert Kaiser, Gordan Krnjaic, Josef Pradler, Andreas Trautner and Sean Tulin for helpful discussions and Saniya Heeba for discovering a typographical mistake in eq.~(\ref{eq:xfint}). This work is supported by the German Science Foundation
(DFG) under the Collaborative Research Center~(SFB) 676 Particles,
Strings and the Early Universe as well as the ERC Starting Grant `NewAve' (638528). 

\appendix

\section{A toy model}
\label{app:toymodel}

In this appendix we discuss how the coupling structure that we consider can arise from a theory with spontaneous breaking of CP symmetry. The starting point is a CP conserving theory that contains a Dirac fermion $\psi$ and a pseudoscalar $P$:
\begin{equation}
 \mathcal{L}_\text{DM} = \bar{\psi} (i \slashed{\partial} - m_0) \psi - i y_\psi P \bar{\psi} \gamma^5 \psi - V(P) \; .
\end{equation}
The fact that the theory is CP-conserving is reflected in the fact that the coupling $y_\psi$ is real and that under a CP transformation $P \rightarrow -P$ so that the Yukawa interaction remains invariant. 

Since $P$ is a SM singlet, it cannot couple directly to either $\bar{q}_L u_R$ or $\bar{q}_L d_R$, both of which are not invariant under $SU(2)$. For a CP-even real scalar singlet $S$ such interactions can be generated after electroweak symmetry breaking (EWSB) via mixing with the SM Higgs. If the scalar potential contains the term $\mu_S H^\dagger H S$, EWSB will induce an off-diagonal mass term of the form $\frac{1}{2} \mu_S \, \vev \, h \, S$, where $\vev$ is the electroweak vacuum expectation value (vev) and $H = \frac{1}{\sqrt{2}} (0, h + \vev)^T$. For a CP-odd real scalar $P$, however, the analogous term $\mu_P H^\dagger H P$ would violate CP and is therefore absent.

To generate couplings of a CP-odd real scalar singlet to SM quarks, one typically assumes the presence of two Higgs doublets, $H_1$ and $H_2$, which contain one physical pseudoscalar degree of freedom $A$ after EWSB~\cite{Ipek:2014gua,No:2015xqa,Goncalves:2016iyg,Bauer:2017ota}. One can then consider a CP-conserving mixing of the form $i \mu_P H_1^\dagger H_2 P + \text{h.c.}$, which after EWSB leads to the mixing term $\frac{1}{2} \vev \, \mu_P \, A \, P$.

The mixing between the pseudoscalar singlet $P$ and the pseudoscalar component of the Higgs doublet then leads to couplings of $P$ to SM fermions $f$:\footnote{We note that in generic two Higgs doublet models it is possible for the pseudoscalar to have e.g.\ enhanced couplings to down-type fermions, depending on the detailed structure of the Yukawa sector and the ratio $\tan \beta$ of the vevs of the two Higgs doublets. We assume here that $\tan \beta \approx 1$ and hence the pseudoscalar couples in the same fashion as the SM-like Higgs boson.}
\begin{equation}
 \mathcal{L}_\text{mixing} \supset - i \sin \theta \sum_f \frac{y_f}{\sqrt{2}} P \bar{f} \gamma^5 f \; ,
\end{equation}
where $y_f$ denotes the SM Yukawa couplings and the mixing angle $\theta$ is given by
\begin{equation}
 \tan 2 \theta = \frac{\mu_P \, \vev}{m_P^2 - m_A^2} \; .
\end{equation}
If $P$ has a mass below the GeV scale, SM precision measurements (for example of rare meson decays) constrain $\sin \theta$ to be very small, typically of order $10^{-4}$ or less (see section~\ref{sec:SMbounds}). Such small mixing angles can easily be achieved if $\mu_P \ll \vev \ll m_A$. In this case the mass eigenstates are almost identical to the interaction eigenstates $P$ and $A$ (and we will hence not make a distinction between the two).

Let us now assume that the potential of the pseudoscalar $P$ is given by
\begin{equation}
V(P) = -\mu^2 P^2 + \lambda_P P^4 \; ,
\end{equation}
so that $P$ obtains a vev $v_P = \mu / \sqrt{2 \lambda_P}$, which spontaneously breaks the CP symmetry. Writing $P = v_P + \phi$, we obtain
\begin{equation}
 \mathcal{L}_\text{DM} = \bar{\psi} \left[i \slashed{\partial} - (m_0 + i y_\psi v_P \gamma^5) \right] \psi - i y_\psi \phi \bar{\psi} \gamma^5 \psi - V(\phi) \; .
\end{equation}
Defining $\tan \alpha = y_\psi v_P / m_0$, we can perform a chiral rotation of the DM field $\psi$ to absorb the complex phase in the DM mass: $\psi \rightarrow \exp(i \gamma^5 \alpha / 2) \psi$:
\begin{equation}
 \mathcal{L}_\text{DM} = \bar{\psi} (i \slashed{\partial} - m_\psi) \psi - y_\psi \phi \bar{\psi} (\cos \delta_\psi + i \sin \delta_\psi \gamma^5) \psi \; ,
\end{equation}
where we have defined $m_\psi = \sqrt{y_\psi^2 v_P^2 + m_0^2}$ and $\delta_\psi = \pi/2 - \alpha$. The fact that $\phi$ obtains both scalar and pseudoscalar couplings to DM makes the spontaneous CP breaking explicit. In fact, for $y_\psi \, v_P \gg m_0$ we find $\delta_\psi \approx 0$, i.e.\ the CP violation in the dark sector is maximal.

In a similar way the spontaneous symmetry breaking will induce a complex mass term for the SM fermions:
\begin{equation}
\mathcal{L}_\text{mass} \supset - \sum_f \bar{f} \frac{y_f}{\sqrt{2}}(\vev + i \sin \theta \, v_P \gamma^5) f \; .
\end{equation}
In complete analogy to the DM field $\psi$ we can now define a phase $\alpha_\text{SM} = \sin \theta \, v_P / \vev$ and perform chiral rotations of the fermion fields to recover real mass terms. It then becomes apparent that the spontaneous CP breaking in the dark sector potentially also induces CP violation in SM observables.

The crucial point is however that as discussed above the mixing angle must be very small, $\sin \theta \lesssim 10^{-4}$, and hence the CP-violating phase is also small: $\alpha_\text{SM} \lesssim 10^{-4} v_P / \vev$.\footnote{We note that another source of CP violation arises from quartic couplings of the form $\lambda_{P1} H_1^\dagger H_1 P^2$, which induce mixing between the SM Higgs and the pseudoscalar singlet once $P$ obtains a vev. We must require that these quartic interactions are sufficiently small that $P$ does not obtain unacceptably large couplings to SM states. In this case, the resulting CP-violating effects will also be small.} In other words, because $P$ is only very weakly coupled to the SM, the spontaneous breaking of CP does not induce any large effects in SM observables. In particular, we can write the couplings of $\phi$ to SM fermions as
\begin{equation}
\mathcal{L}_\text{mixing} = - y_\text{SM} \sum_f \frac{y_f}{\sqrt{2}} \phi \bar{f} (\cos \delta_\text{SM} + i \sin \delta_\text{SM} \gamma^5) f
\end{equation}
with $y_\text{SM} = \sin \theta$ and $\delta_\text{SM} = \pi / 2 - \alpha_\text{SM}$. We then find $\cos \delta_\text{SM} \approx \alpha_\text{SM} \ll 1$ and $\sin \delta_\text{SM} \approx 1$, i.e.\ $\phi$ has almost exclusively pseudoscalar couplings to SM states.

To conclude this discussion, let us briefly review experimental bounds on light spin-0 bosons with CP-violating couplings. For $m_\phi < m_e$ such a light scalar can potentially induce a sizeable electric dipole moment of the electron, which is constrained to be \mbox{$|d_e| < 8.7 \cdot 10^{-29} \, e \, \mathrm{cm}$}~\cite{Baron:2013eja}. The one-loop contribution is given by~\cite{Chen:2015vqy}
\begin{equation}
d_e = y_\text{SM}^2 \frac{m_e}{\vev^2}\frac{e}{16\pi^2} \sin 2\delta_\text{SM} \; ,
\end{equation}
whereas the two-loop Barr-Zee contribution is approximately given by~\cite{Marciano:2016yhf}
\begin{equation}
d_e \sim y_\text{SM}^2 \frac{m_e}{\vev^2} \frac{e}{16 \pi^2} \frac{\alpha}{\pi} \sin 2\delta_\text{SM} \mathcal{F}(m_\phi) \log (\Lambda / m_\phi) \; . 
\end{equation}
Here $\mathcal{F}(m_\phi)$ denotes the form factor for the effective vertex $\phi F^{\mu\nu} \tilde{F}_{\mu\nu}$ obtained from integrating out heavy quarks, mesons and leptons. For $m_\phi \sim m_e$ and $\Lambda \sim 1\:\text{TeV}$, one obtains $\mathcal{F}(m_\phi) \log (\Lambda / m_\phi) \sim 10^2$, meaning that the one-loop and two-loop contributions are of similar magnitude.\footnote{Higher-order contributions from light-by-light scattering and vacuum polarisation are found to be sub-dominant~\cite{Marciano:2016yhf}.} The experimental bounds is thus satisfied for $y_\text{SM}^2 \sin 2\delta_\text{SM} \lesssim 10^{-4}$, which is always the case in the parameter region that we consider. Experimental bounds on the electric dipole momenta of the muon, the neutron or mercury nuclei give comparable or weaker constraints.

\section{Dark matter self-interactions from pseudoscalar exchange}
\label{app:pseudoSIDM}

\subsection*{Tree-level analysis}

The self-interactions of DM particles induced by the exchange of a spin-0 boson $\phi$ can be calculated by solving the Schr\"odinger equation for $\psi$ in the presence of a non-relativistic scattering potential $V(r)$. This potential is obtained by taking the Fourier transform of the matrix element $\mathcal{M}_{\psi \psi \rightarrow \psi \psi}$ with respect to the exchanged three-momentum $\vec{q}$. For the scalar coupling $\bar \psi \psi \phi$ between DM and the mediator, at lowest order in $\vec{q}^2$ this procedure gives rise to an attractive Yukawa potential,
\begin{align}
V_{\text{S}}(r) = -\alpha_\text{S} \frac{e^{-m_\phi r}}{r}
\end{align}
with $\alpha_\text{S} \equiv (y_\psi \cos \delta_\psi)^2/(4 \pi)$, which can induce strong self-interactions of DM. For the pseudoscalar coupling $\bar \psi \gamma^5 \psi \phi$, at lowest non-vanishing order in $\vec{q}^2$ one obtains
\begin{align}
V_{\text{PS}}(r) = &\frac{\alpha_\text{PS}}{12} \frac{m_\phi^2}{m_\psi^2} \left( \frac{e^{-m_\phi r}}{r} - \frac{4 \pi}{m_\phi^2} \delta^{(3)} (\vec{r}) \right) \vec{\sigma}_1 \cdot \vec{\sigma}_2 \nonumber \\
&+ \frac{\alpha_\text{PS}}{12}  \frac{m_\phi^2}{m_\psi^2} \left( 1 + \frac{3}{m_\phi r} + \frac{3}{m_\phi^2 r^2} \right) \frac{e^{-m_\phi r}}{r} \, S_{12} (\vec{r}) \,,
\label{eq:VPS}
\end{align}
with $\alpha_\text{PS} \equiv (y_\psi \sin \delta_\psi)^2/(4 \pi)$ and $S_{12} (\vec{r}) \equiv 3 (\vec{\sigma}_1 \cdot \hat{r})( \vec{\sigma}_2 \cdot \hat{r}) - \vec{\sigma}_1 \cdot \vec{\sigma}_2$, which agrees with~\cite{Daido:2017hsl}, including the overall sign. The Yukawa-like part of this potential is suppressed by $m_\phi^2/m_\psi^2$, and hence is irrelevant for the self-scattering of DM. The part proportional to $1/r^3$, on the other hand, is not suppressed in the limit $m_\phi \ll m_\psi$, and could potentially lead to strong self-interactions. However, it is well-known that the singular behaviour of a $1/r^3$ potential leads to ill-defined solutions of the Schr\"odinger equation; in particular a naive calculation of the scattering amplitude for DM self-scattering diverges, making it impossible to directly extract information about the self-scattering cross section of DM~\cite{Bedaque:2009ri,Bellazzini:2013foa}.

It is however by no means clear that the $1/r^3$ singularity is actually physical. The potential given by eq.~(\ref{eq:VPS}) is derived by taking the Fourier transform of the matrix element $\mathcal{M}_{\psi \psi \rightarrow \psi \psi}$ and expand the resulting expression for small relative momenta $\vec{q}$, corresponding to large values of $r$. Hence, to investigate the behaviour of $V(r)$ for $r \rightarrow 0$ more and more powers of $q$ have to be taken into account, leading to additional terms $1/r^n$ in the potential, which are expected to regulate the behaviour of the potential at the origin.

In fact, exactly the same argument applies also to the simpler case of scalar or vector exchange. For example, the famous term leading to the spin-orbit coupling in the hydrogen atom is given by $V_\text{spin-orbit} \propto (\vec{L} \cdot \vec{S})/r^3$. In the familiar context of the fine splitting of the hydrogen energy  levels, the singularity of this potential is not a problem, as in first order perturbation theory one is only interested in the finite expectation value of this operator with respect to the unperturbed wave function. However, one would encounter the same problems with the singular behaviour at $r \rightarrow 0$ when attempting to actually solve the Schr\"odinger equation for the hydrogen atom including this correction term. In other words, when worrying about the singular behaviour of the pseudoscalar potential $V_{\text{PS}}(r)$, one should also worry about the corresponding singularities in the scalar potential $V_{\text{S}}(r)$. Intuitively, the higher-order terms are then expected to give rise to only subdominant contributions to the scattering cross section of DM, suggesting that the exchange of a pseudoscalar does not give rise to strong self-interactions of DM compared to the cross section induced by scalar exchange. Hence, in the following we will only take into account the scattering of DM off the lowest-order potential $V_{\text{S}}(r)$ given by eq.~(\ref{eq:VS}).\footnote{For the same reason we also neglect the monopole-dipole potential which is present for CP-violating phases $\delta_\psi$~\cite{Daido:2017hsl}: All terms in that potential are suppressed either by $m_\phi$ or by additional powers of $1/r$.}

\subsection*{Possible impact of one-loop corrections}

It has been proposed that the exchange of two pseudoscalars via a box diagram could effectively give rise to a scalar coupling between the DM particles and hence to larger self-interaction rates~\cite{Oshima}. We have repeated the calculation of these loop diagrams, and we find that the momentum and velocity dependence of the resulting scattering amplitude \emph{cannot} be described by the effective exchange of a scalar particle. In particular, in the limit $v \rightarrow 0$ the scattering cross section is given by
\begin{align}
\sigma_{\psi \psi \rightarrow \psi \psi}^{\text{(box)}} = \frac{\alpha_\text{PS}^4}{64 \pi^2 m_\psi^2} \left[ -3 + \log \left( \frac{m_\psi^2}{m_\phi^2} \right) \right]^2
\end{align}
with $\alpha_\text{PS}$ defined as above. This is drastically different from the tree-level exchange of a scalar particle, which gives rise to a cross section $\propto 1/m_{\text{scalar}}^4$ in the limit $v \rightarrow 0$. In particular, we find numerically that the one-loop induced cross section is always significantly too small in order to give rise to relevant self-interactions of DM particles. This finding agrees with similar conclusions obtained for loop-induced direct detection cross sections~\cite{Freytsis:2010ne}.

\section{Dark matter self-interactions from scalar exchange}
\label{app:selfi}

In this appendix we summarize the formalism for the calculation of the momentum transfer cross section $\sigma_\mathrm{T}$ arising from the scalar coupling of the DM particle to the spin-0 mediator $\phi$. In contrast to previous works, we fully take into account effects arising from the indistinguishability of the scattered particles. Under the assumption that there is no asymmetry in the abundances of $\psi$ and $\bar \psi$, the averaged momentum transfer cross section is given by
\begin{equation}
\sigma_\mathrm{T} = \frac12 \left( \sigma_\mathrm{T}^\text{PP} + \sigma_\mathrm{T}^\text{PA}\right) \; ,
\end{equation}
where PP (PA) denotes particle-particle (particle-antiparticle) scattering and
\begin{equation}
\sigma_\mathrm{T}^\text{PP,PA} = \int \text{d} \Omega \, (1 - |\cos \theta |)\left( \frac{\text{d}\sigma}{\text{d} \Omega} \right)^\text{PP,PA} \;.
\label{eq:sigmaT_definition_general}
\end{equation}
The factor $1/2$ for two identical particles in the final state has been included implicitly in the definition of $\left( \text{d}\sigma / \text{d} \Omega \right)^\text{PP}$.

\subsubsection*{Born regime}

Non-perturbative effects in the scattering process are negligible as long as $\alpha_S m_\psi / m_\phi \ll 1$ (the Born regime), where $\alpha_S \equiv y_\psi^2 \cos^2 \delta_\psi / (4 \pi)$ denotes the coupling strength relevant for DM self-interactions. We find
\begin{align}
\sigma_\mathrm{T}^\text{PP} \big|_\text{Born} &= \frac{4 \pi \alpha_S^2}{m_\psi^2 v^4} \left[ 6 \log \left(\frac{m_\psi^2 v^2}{2 m_\phi^2} +1\right) - \frac{4 m_\psi^2 v^2+ 6 m_\phi^2}{m_\psi^2 v^2 +2 m_\phi^2} \log\left(\frac{m_\psi^2 v^2}{m_\phi^2} + 1\right) \right] \; , \\
\sigma_\mathrm{T}^\text{PA}\big|_\text{Born} &=  \frac{8 \pi \alpha_S^2}{m_\psi^2 v^4}  \log\left(\frac{(m_\psi^2 v^2 + 2 m_\phi^2)^2}{4 m_\phi^2 (m_\psi^2 v^2 + m_\phi^2)}\right) \; .
\end{align}
Compared to~\cite{Tulin:2013teo}, where only the $t$-channel contribution is included and the momentum transfer cross section is defined via $\int \text{d} \Omega \, (1 - \cos \theta)\left( \text{d}\sigma/\text{d} \Omega \right)$, our full calculation in the Born regime gives the same result in the limit $m_\psi v \gg m_\phi$ both for particle-particle and particle-antiparticle scattering. On the other hand, for $m_\psi v \ll m_\phi$ the expressions above are smaller by a factor of 4 (2) for particle-particle (particle-antiparticle) scattering compared to the results obtained in~\cite{Tulin:2013teo}.

\subsubsection*{Non-perturbative regime}

For $\alpha_S m_\psi / m_\phi \gtrsim 1$, non-perturbative effects corresponding to the multiple exchange of the mediator $\phi$ have to be taken into account. In practice, one numerically solves the Schr\"odinger equation corresponding to the scattering potential using the standard techniques of partial wave decomposition (see also appendix~\ref{app:pseudoSIDM}). The scattering amplitude can be written as
\begin{align}
f(\theta) = \frac{2}{m_\psi v} \sum_{l=0}^\infty (2l+1) e^{i \delta_l} \sin \delta_l P_l (\cos \theta) \,,
\label{eq:f_theta}
\end{align}
where we obtain the phase shifts $\delta_l$ by employing the numerical technique described in~\cite{Tulin:2013teo}. For scattering of identical particles, the differential scattering cross section then follows from
\begin{align}
\left( \frac{\text{d}\sigma}{\text{d} \Omega} \right)^\text{PP}_\xi = \left| f(\theta) + \xi f(\pi - \theta)\right|^2
\end{align}
with $\xi = +1$ ($-1$) if the spatial wave function is symmetric (antisymmetric) under particle exchange. If scattering occurs via the spin-singlet channel, the spin wave function is antisymmetric, and hence the spatial wave function has to be symmetric; correspondingly, for scattering via the spin-triplet channel the spatial wave function is antisymmetric. Assuming unpolarized DM particles, we obtain
\begin{align}
\left( \frac{\text{d}\sigma}{\text{d} \Omega} \right)^\text{PP} = \frac14 \left( \frac{\text{d}\sigma}{\text{d} \Omega} \right)^\text{PP}_{\xi = +1} + \frac34 \left( \frac{\text{d}\sigma}{\text{d} \Omega} \right)^\text{PP}_{\xi = -1}
\end{align}
On the other hand, since particle and anti-particle are distinguishable, one simply obtains
\begin{align}
\left( \frac{\text{d}\sigma}{\text{d} \Omega} \right)^\text{PA} = \left| f(\theta) \right|^2 \,.
\end{align}
The momentum transfer cross sections $\sigma_\mathrm{T}$ then follows from eq.~(\ref{eq:sigmaT_definition_general}) by numerically integrating over the angular variable $\theta$.\footnote{Due to the presence of the weighting factor $1 - |\cos \theta|$ instead of $1 - \cos \theta$ in the definition of $\sigma_\mathrm{T}$, there is no simple analytical expression for the momentum transfer cross section in terms of the phase shifts $\delta_l$.} 

\subsubsection*{Classical regime}

\begin{figure}
\begin{center}
\includegraphics[scale=1.2]{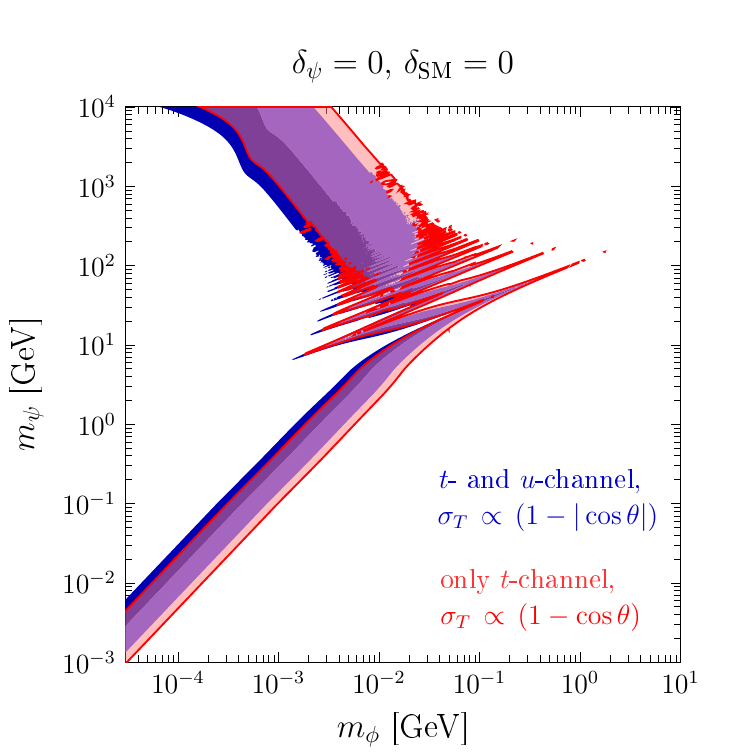}
\end{center}
\caption{Regions of the parameter space leading to sufficiently large DM self-interactions on dwarf scales, obtained via the full calculation of $\sigma_T$ (blue shaded bands) and via the approximation of classical distinguishability in the scattering process (red curves).}
\label{fig:comparison}
\end{figure}

For $m_\psi v \gg m_\phi$ (denoted as classical regime), more and more partial waves $l$ have to be taken into account in the calculation of the scattering amplitude in eq.~(\ref{eq:f_theta}), in order to obtain a sufficiently convergent series. At some point (in our case typically at $l_\text{max} \sim 100$), the numerical approach becomes infeasible, and instead we use an approach based on the fitting functions for $\sigma_\mathrm{T}$ provided in~\cite{Cyr-Racine:2015ihg}. However, these results have been obtained using only $t$-channel exchange and adopting the definition $\int \text{d} \Omega \, (1 - \cos \theta )\left( \text{d}\sigma/\text{d} \Omega \right)$ for the momentum transfer cross section. We find that in order to match our results based on the numerical solution of the Schr\"odinger equation sufficiently smoothly onto the fitting functions in the regime where $m_\psi v \gg m_\phi$, we need to multiply the expressions given in~\cite{Cyr-Racine:2015ihg} by a factor $1/2$. This factor of $1/2$ can be understood from the fact that $\text{d}\sigma/\text{d}\Omega$ is approximately independent of $\cos \theta $ in the classical regime~\cite{Tulin:2013teo} and that
\begin{equation}
\int_{-1}^1 \mathrm{d} \cos\theta (1 - |\cos \theta|) = \frac{1}{2} \int_{-1}^1 \mathrm{d}\cos \theta (1 - \cos \theta) \; .
\end{equation}

To conclude this discussion, we illustrate in figure~\ref{fig:comparison} the impact of using the correct definition for the momentum transfer cross section. The blue regions correspond to the full calculation, which takes into account the indistinguishability of two DM particles. The red lines, on the other hand, illustrate the results that one obtains from the simpler calculation, including only $t$-channel scattering and defining the momentum transfer cross section via $\int \text{d} \Omega \, (1 - \cos \theta ) \text{d}\sigma/\text{d}\Omega$.

\providecommand{\href}[2]{#2}\begingroup\raggedright\endgroup

\end{document}